\documentclass[12pt]{iopart} \usepackage{graphicx} \usepackage{iopams}
\usepackage{amsmath} \usepackage{bm} \def\msum{\displaystyle\sum}
\begin{document}

\title{Self-organization without conservation : \\ true or just apparent
  scale-invariance?}

\author{Juan A. Bonachela and Miguel A. Mu\~noz  }

 \address{Departamento
  de Electromagnetismo y F{\'\i}sica de la Materia and \\ Instituto de
  F{\'\i}sica Te{\'o}rica y Computacional Carlos I, \\ Facultad de
  Ciencias, Universidad de Granada, 18071 Granada, Spain} 

 \date{\today}

\begin{abstract} 
  The existence of true scale-invariance in slowly driven models of
  self-organized criticality {\it without} a conservation law, as
  forest-fires or earthquake automata, is scrutinized in this paper.
  By using three different levels of description - (i) a simple mean
  field, (ii) a more detailed mean-field description in terms of a
  (self-organized) branching processes, and (iii) a full stochastic
  representation in terms of a Langevin equation-, it is shown on
  general grounds that non-conserving dynamics does {\it not} lead to
  {\it bona fide} criticality.  Contrarily to conserving systems, a
  parameter, which we term ``re-charging'' rate (e.g. the tree-growth
  rate in forest-fire models), needs to be {\it fine-tuned} in
  non-conserving systems to obtain criticality.  In the infinite size
  limit, such a fine-tuning of the loading rate is easy to achieve, as
  it emerges by imposing a second separation of time-scales but, for
  any finite size, a precise tuning is required to achieve criticality
  and a coherent finite-size scaling picture.  Using the approaches
  above, we shed light on the common mechanisms by which ``apparent
  criticality'' is observed in non-conserving systems, and explain in
  detail (both qualitatively and quantitatively) the difference with
  respect to true criticality obtained in conserving systems. We
  propose to call this {\it self-organized quasi-criticality} (SOqC).
  Some of the reported results are already known and some of them are
  new. We hope the unified framework presented here helps to elucidate
  the confusing and contradictory literature in this field.  In a
  second accompanying paper, we shall discuss the implications of the
  general results obtained here for models of neural avalanches in
  Neuroscience for which self-organized scale-invariance in the
  absence of conservation has been claimed.
\end{abstract}

\vspace{2pc}

\submitto{Journal of Statistical Mechanics}

\maketitle

\section{Introduction: Critical Self-organization with and without
  conservation}
\label{Intro}

Power-law distributions are quite common in Nature: earthquakes and
micro-fractures, solar flares, weather records, snow avalanches, crackling,
and $1/f$ noise are just a few examples of systems displaying scale-invariance
\cite{Mandelbrot}. Many of these, among many others, have been claimed to be
critical, i.e.  to lie at, or very close to, a critical point, which ensues
scale-invariance and the concomitant power-law distributions.  It is worth
stressing that power-laws (or approximate power-laws) emerging from the
complex interactions of many-component systems placed at the vicinity of a
critical point and, therefore, with diverging correlation lengths and the
associated power-law decay of temporal and spatial correlations, should be
clearly distinguished from other power-law distributions arising in many
different contexts (word distributions, city populations, citations,
etc). These latter can be generated by a wealth of {\it non-critical
  mechanisms} as, for instance, multiplicative noise or fragmentation
processes (see \cite{Sornette,Mitzen,Newman} for recent reviews), without the
need to invoke criticality.

For a system to be critical, it does not suffice to have power-law
distributed observables but, more crucially, it has to obey {\bf
  finite-size scaling}: measurements at various system-sizes (scales)
can be related to each other by re-scaling variables and quantities in
some specific ``scale-invariant'' way (i.e. scaling collapses can be
performed).

Given that, in standard phase transitions (both in equilibrium and
away from it), a precise parameter tuning is required to reach
criticality and generate power-law distributed quantities, an
alternative explanation for the emergence of generic critical
scale-invariance (i.e.  criticality occurring without requiring
fine-tuning) was historically much needed \cite{generic}. How does
criticality emerge spontaneously?

In a seminal paper, Bak, Tang and Wiesenfeld (BTW) \cite{BTW} introduced, back
in $1987$, the concept of {\bf self-organized criticality} (SOC)
\cite{Bak,Jensen,Moloney,Dhar,Turcotte,Frigg} aimed at solving the previous
conundrum. The research line opened by their breakthrough work continues to
attract, more than twenty years after, a great deal of interest.  From this
perspective, the overwhelming activity generated in the last decade around
{\it scale-free networks} \cite{BA} can be certainly considered as a prominent
extension of preceding research on SOC.

A handful of mechanisms were proposed under the common name of SOC to
justify the abundant presence of scale-invariance in the natural
world.  The most successful among them is the one exemplified by the
original BTW {\bf sandpile model}.  Many variations of the BTW
sandpile were proposed: the Manna sandpile \cite{Manna}, the Oslo
rice-pile model \cite{Oslo}, and the Zhang model \cite{Zhang}, are
just a few of them.  For the sake of completeness and for future
reference, a brief description of the best-known prototypical SOC
models is provided in Appendix A.  Other possible mechanisms, as {\it
  extremal dynamics} \cite{extremal}, have been studied, but we shall
not be concerned with them here.

The idea inspiring sandpiles, ricepiles, and related SOC models is
that many systems in Nature, when pushed/driven slowly, respond very
irregularly with rapid re-arrangements (avalanches, bursts) of a broad
variety of sizes. The distribution of such avalanches is
scale-invariant in many cases. Of course, sandpile SOC models are too
simplistic to reproduce the detailed behavior of real sand-piles; they
can be regarded as ''metaphors'' capturing only some particular
features of real systems in a stylized manner. Nevertheless, after
many (partially) failed attempts \cite{exp-no} power-laws for
avalanche in real granular piles, obeying finite-size scaling, were
experimentally measured \cite{exp-si}. Moreover, other physical
situations, as vortex avalanches in type-II superconductors, can be
mimicked as sandpiles, and their critical properties rationalized in
terms of these \cite{superc}.

The dynamics of a generic sandpile model can be synthesized as follows: some
type of ``energy'' or ``stress'' (sand-grains) is progressively injected in
discrete units (i.e. grains are dropped singly) at the sites of a spatially
extended system (usually a two-dimensional square lattice) at a slow
timescale.  Whenever a certain threshold of local energy is overcome, the
corresponding site becomes unstable and its accumulated energy is
redistributed at a much faster timescale among its neighbor sites.  These, on
their turn, can become unstable, and trigger a cascade of re-arrangements,
i.e.  an {\bf avalanche} or outburst of activity.  Local redistribution rules
are conserving in sandpiles: energy does not dissappear, but only diffuses
around. Open boundaries are customarily considered to allow for energy release
from the system.  Once all activity ceases (i.e.  the avalanche stops) new
energy is injected (i.e. a grain is dropped) into the system, and so on, until
a statistically stationary state is reached.  In such a steady state
avalanches are scale invariant, i.e their sizes/times are power-law
distributed up to a maximum scale imposed by the system size: the system
``self-organizes'' to a critical point
\cite{Bak,Jensen,Moloney,Dhar,Turcotte,GG,BJP,Mikko}.  Moreover, the
associated power spectrum exhibit $1/f^{\alpha}$ noise \cite{Laurson}.

For the sake of generality, let us fix a nomenclature common to all models
discussed in this paper: ``{\bf energy}'' refers to the accumulated and
transported magnitude and ``{\bf activity}'' describes ``energy above
threshold''. In some cases, sites below threshold will be sub-classified in
two groups: {\it critical}, which can become active upon receiving one input
of energy (for instance, one grain) and {\it stable}, which cannot \cite{MF}.

An essential ingredient of SOC is {\bf slow driving} (driving and
dynamics operating at two infinitely separated timescales
\cite{Bak,Jensen,Moloney,GG,Fly}, i.e. avalanches are instantaneous
relative to the timescale of driving).  Such an infinite separation is
usually achieved by driving the system only when all activity has
stopped, but not during avalanches.  For any {\it finite} separation
of timescales, a finite characteristic (time/size) scale appears;
hence, slow driving is a crucial requirement for generic
scale-invariance to emerge \cite{GG,GG2,GG3,finite}.

It was soon emphasized that {\bf energy conservation} is also a key
player for criticality to emerge in sandpile models
\cite{GG,Bak-non,Hwa}.  Note that by ``conserving system'' one can
refer either to models with bulk conserving dynamics and boundary
dissipation, or to cases with a bulk dissipation rate vanishing in the
large system size limit \cite{Malcai}. Various relatively simple
arguments were proposed to rationalize the existence of true
criticality in the steady state of {\it conserving} self-organized
systems (some of them are briefly discussed in the next section).
However, these arguments cannot be easily extended to similar
non-conserving systems.  In particular, different works have shown
that the level of dissipation acts as a relevant parameter in the
renormalization group sense: any degree of bulk dissipation breaks
criticality in sandpile models \cite{GG,Hwa,non}.

Still, given the large variety of natural phenomena exhibiting (exact
or approximate) scale-invariance in which some form of dissipation is
inevitably present (i.e. systems without any obvious conserved
quantity), alternative mechanisms for self-organization to criticality
in the absence of conservation were needed to achieve a comprehensive
picture of generic scale-invariance \cite{GG}.

Two acclaimed non-conserving self-organized models, or better, two {\it
  families of models} proposed to fill the gap between theoretical
understanding and empirical facts are earthquake and forest-fire models (see
Appendix A for definitions). These are highly non-trivial, and interesting
models with rich and complex phenomenology.  Owing to the lack of solid
theoretical arguments, analogous to the ones sketched above for conserving
systems, and despite of numerical evidence showing power-laws for some
decades, the existence of true generic scale-invariance in them has been long
controversial.

It is beyond the scope of this paper to review exhaustively the large body of
interesting literature devoted to non-conserving SOC models, some aspects of
which remain unsettled. But, let us just underline that the state-of-the-art
is, as documented in the next section, that {\it none of the considered
  non-conserving models is truly critical; they just show ``apparent
  scale-invariance'' or ``dirty criticality'' for some decades}.

However, this final conclusion has not been sufficiently stressed and it has
certainly not permeated the literature. This is likely due to the absence of a
general theory, which may suggest that the results discussed above are
specific to each particular model. Indeed, works continue to be published
assuming or claiming true criticality for SOC non-conserving systems.  For
instance, in \cite{PJ} an interesting and solvable ``non-conserving model of
SOC'' was proposed and studied analytically. In a more recent series of
papers, Juanico and collaborators claim to have constructed different
non-conserving self-organizing models with applications in various fields (as
neuroscience, population dynamics, etc.)  \cite{Juanico}.  Also, in a recent
work, Levina {\it et al.}  propose a non-conserving SOC model for {\it neural
  avalanches} to capture the apparent scale-free behavior of avalanches of
activity observed experimentally in networks of cortical neurons
\cite{Levina}.  An exhaustive analysis of this last model, as well as a study
of the possible relation between SOC and neural avalanches, is left for a
separate publication.

The aim of the present paper is to put together some previously existing
results, scattered in the literature, (although this is not intended to be an
exhaustive review article) and, more importantly, to rationalize the
conclusion that none of the above mentioned non-conserving models, nor
variations of them, exhibits true criticality, within a {\it unified
  framework}. To this end, we rely on different types of analytical arguments
complemented by computer simulations. In passing, we shall report on a number
of new results and present a critical discussion on the existence of true
scale-invariance in Nature.

The rest of the paper is structured as follows: in Section \ref{uno}, we
briefly review conserving and non-conserving models of SOC, as well as some
arguments to justify the existence of criticality in the first group. In the
remaining Sections, we elucidate the existence or not of criticality in
non-conserving systems using different approaches of increasing complexity. In
particular, in Section \ref{zeroth}, we discuss a simple mean-field approach
based on an energy balance equation; it is useful to illustrate some key
concepts as the ``loading mechanism''.  In Section \ref{SOBP}, we study a
self-consistent mean-field approximation, namely the so-called {\it
  self-organized branching process}; it serves as an adequate benchmark to
scrutinize the effects of dissipation and ``loading'' in critical
self-organization.  In Section \ref{Langevin}, we present (and briefly review)
the Langevin theory of conserving SOC systems. It constitutes a solid basis to
implement dissipation and loading in a systematic way and to provide clear
evidence on the lack of criticality in non-conserving models. Finally, the
conclusions and a critical discussion of the implications of our main results
are presented in Section \ref{Last}.

\section{Conserving versus non-conserving models of SOC}
\label{uno}

\subsection{Conservation and criticality.}

As said above, different type of arguments of different nature justify the
existence of true criticality in conserving SOC models. Some of them are as
follows:

\begin{itemize} 
\item 

  The energy, introduced into a pile at generic sites, can reach the
  boundaries (and, thus, be dissipated) only by means of the diffusive
  transport of grains occurring during avalanches. Owing to this, and
  provided that a steady state exists, arbitrarily large avalanches
  (of all possible sizes) should exist for an arbitrarily large system
  size, ensuing a power-law size-distribution.  Contrarily, in the
  presence of non-vanishing bulk-dissipation, energy disappears at
  some finite rate, and avalanches stop after some characteristic
  lifetime/size determined by the dissipation rate \cite{GG}.

  This type of argument, even if commonly used in the literature, is
  (at best) incomplete, and can be misleading.  It does not consider
  the possibility of having a characteristic scale larger than the
  system size, which would allow for avalanches to reach the
  boundaries.  Actually, this is what happens in many real sandpiles
  (with inertial effects): energy is dissipated quasi-periodically in
  large system-wide avalanches rather than in scale-invariant
  avalanches (see the chapter on experimental set-ups of sandpiles in
  \cite{Jensen,Moloney,Tesis} or \cite{GG}).

\item From a more abstract viewpoint (not referring specifically to
  sandpiles or SOC), energy conservation follows from the existence of
  a continuous symmetry (in this case, {\it temporal translational
    invariance}) as a consequence of Noether's theorem \cite{Noether}.
  This also holds the other way around: time translational invariance
  implies energy conservation.  When energy conservation is violated,
  the corresponding symmetry is broken and a characteristic (finite)
  time-scale appears generically.

\item From a field-theoretical perspective, in order to have scale-invariance,
  generic infrared divergences are required. But these are generically lost in
  the presence of a non-vanishing linear ``mass'' term (adopting the
  field-theory jargon) as, for instance, a dissipative term.  In particular,
  if a term $-\epsilon\psi$, is introduced into the simplest mesoscopic
  equation for a diffusive field $\psi(\vec{x},t)$
\begin{equation}
  \partial_{t}\psi(\vec{x},t)=D_{\psi}\nabla^{2}\psi(\vec{x},t)+\eta(\vec{x},t),
\label{lang_cons}
\end{equation}
where $\eta(\vec{x},t) $ is a zero-mean Gaussian white noise, a simple
calculation reveals that the equal time two-point correlation function can be
written as
\begin{equation}
\langle \psi(x,t) \psi(0,t)\rangle \sim \frac{\exp{(-x/\xi)}}{x^{(d-1)/2}},
\label{correlator}
\end{equation}
for $x$ much larger than the correlation length $\xi \sim
1/\sqrt{\epsilon}$. Accordingly, it is only for $\epsilon =0$ that $\xi$
diverges, the exponential cut-off disappears, and the correlation function
decays algebraically; for any non-vanishing value of $\epsilon$ there is a
{\it size-independent} exponential cut-off.  Something similar occurs for
other correlation functions.

Note that the noise in Eq.(\ref{lang_cons}) is not fully conserving, but only
conserving on average (i.e. conservation needs only to hold on average to
preserve scale invariance \cite{Kertesz}). The same conclusions can be also
deduced for an equation analogous to Eq.(\ref{lang_cons}) but with a strictly
conserving noise (see \cite{GG,GG2,GG3}).

\item Last but not least, a mesoscopic Langevin equation that captures
  the critical properties of stochastic sandpiles and related models
  with a conservation law has been proposed \cite{BJP,FES0,FES,FESJabo}. It
  describes systems with many absorbing states (which correspond to
  the many stable microscopic configurations of a sandpile) and a
  conservation law. While a detailed description of this is left for a
  forthcoming section, we just stress here that it constitutes a sound
  field theoretical representation of conserving SOC, reproducing all
  critical exponents. The underlying mechanism of SOC, highlighted by
  this theory, cannot be straightforwardly generalized to
  non-conserving systems without including an additional fine tuning
  (as we shall show in Section \ref{Langevin}).

\end{itemize}

In summary, {\it there exists solid theoretical ground to underpin the
  existence of true criticality in conserving self-organized systems.
  The same type of arguments cannot be easily extended to
  non-conserving systems}.  In particular, as said already,
non-conservative sandpiles have been explicitly shown to be
non-critical.

\subsection{Critical self-organization without conservation?}

The two main prototypical non-conserving ``self-organized'' models (or
families of models), studied profusely in the literature, are (see
Appendix A):

\begin{itemize} 
\item {\bf Earthquake models} as the Olami-Feder-Christensen (OFC) \cite{OFC}
  and related ones \cite{BK,Carlson,QBak,QFeder,Christensen}, in which the
  degree of dissipation can be explicitely controlled, and for which
  conservation is preserved only for some specific choice of a parameter.

\item {\bf Forest fire} models as the Drossel-Schwabl automaton
  \cite{FFM2} (and earlier versions of it \cite{oldFFM}), in which
  there is no conserved quantity whatsoever.
\end{itemize}

Together with bulk-dissipation, these two models have a common key ingredient,
absent in conserving systems: there is an increase of the ``background
energy'' at some (or at all) sites, occurring between avalanches; {\it (i)}
the accumulated stress at each site grows continuously between quakes in
earthquake models as the OFC, and {\it (ii)} new trees grow between two
consecutive fires in forest-fire models (see below).

The effect of these ``{\bf loading mechanisms}'', as we shall call them
generically, is to counterbalance the loss of ``energy'' (grains, stress,
trees) produced by dissipation and, in this manner, try to restore
conservation on average and, thus, criticality \cite{MF,PJ,Juanico}.  However,
let us caution that such a compensation needs to be exact and, therefore,
unless a new mechanism giving rise to a perfect cancellation is devised,
fine-tuning of the loading rate is the only obvious way in which conservation
can be restored.

 Let us now discuss in more detail these non-conserving archetypical models.

\subsubsection{Earthquakes:}

The Olami-Feder-Christensen cellular automaton \cite{OFC} is a simplified
version of a previously proposed fault dynamics models: the spring-block of
Burridge-Knopoff model \cite{BK} and related {\it stick-slip models}
\cite{QBak,QFeder} designed to capture the essence of earthquakes (the
Gutenberg-Richter law \cite{Gutenberg} for the distribution of magnitudes) as
well as of similar systems with friction and jerky motion (see Appendix A).

At each time step, the ``forces'' $F(i,j)$ (or energies, to stick to our
generic terminology), defined at each site of a two-dimensional lattice, are
increased at a constant rate.  Whenever $F$ at any site reaches the threshold
value, $F_{thr}=1$, it is reset to zero and the forces at its $4$ nearest
neighbors are increased by an amount $\delta F= \alpha F(i,j)$. This might
trigger cascades of re-arrangements, i.e. avalanches.  Observe that the bulk
dynamics is conserving only for $\alpha=1/4$ in two-dimensions.

Early computer simulations and theoretical results
\cite{OFC,LiseJensen,Janosi} seemed to support the existence of
criticality for values of $\alpha$ as low as $\alpha \approx 0.05$.
It was also early reported that, imposing periodic boundary
conditions, the OFC model enters a cycle of periodic configurations
with no sign of criticality whatsoever \cite{QGrassberger,GG3,QMT}.
This suggests that the bulk dynamics is profoundly influenced by
boundary conditions.  Actually, it was proposed that boundary-induced
heterogeneity is essential to obtain partial synchronization between
clusters of different sizes and that such a partial {\it
  synchronization or ``phase-locking'' mechanism} is at the basis of
the OFC complex behavior \cite{QGrassberger,GG3,QMT}.

The role of different features (as, for instance, changes in the boundary
conditions, introduction of quenched disorder in the local rules, lattice
topology, etc.) on synchronization and their effects in the properties of the
OFC model have been largely analyzed in the literature
\cite{LiseJensen,Bhatta,Ghaffari,Mousseau,Kinouchi,Caruso2,Lise,Kotani,Bottani}.
It has been also shown that results are affected by numerical precision
\cite{precision}.  The overall picture is that the ``synchronization
mechanism'', even if fascinating, is too fragile as to be a solid explanation
for generic emergence of criticality.  A nice and rather exhaustive review of
the literature on the OFC model and variations of it can be found in
\cite{WD}.

On the analytical side, Br\"oker and Grassberger \cite{Broker} and Chabanol
and Hakim \cite{Hakim}, in two independent papers, were able to calculate the
energy distribution, the effective branching ratio, and the average avalanche
size for a {\it random-neighbor} version of the OFC model, which turns out to
be analytically solvable. Their main conclusion is that it is only in the
conserving limit that the model becomes critical, while exponential cut-offs
appear for any $\alpha \neq 1/4$.  Similarly, de Carvalho and Prado claimed,
relying on an effective branching ratio analysis \cite{Prado}, that the OFC
model is only critical in the conserving limit (see also \cite{Miller}).

Remarkably, it is also shown in \cite{Broker} and \cite{Hakim} that the
average avalanche size is distributed as a power-law with a cut-off function,
$\exp{(\frac{4}{1-4\alpha})}$, which diverges in a very fast way when the
conserving limit is approached.  This provides an explanation for the relative
large power-law regimes observed even in the non-conserving case. It would be
certainly nice to have extensions of this result to other non-mean-field like
systems.

In a similar line of reasoning, Kinouchi and Prado introduced the
concept of ``{\it robust criticality}'' or ``{\it almost
  criticality}'' \cite{Kinouchi2}: for a fixed dissipation rate,
systems with a loading mechanism are closer to criticality than
systems without it. The reason for this is simple: moderate
loading partially compensates energy dissipation.

Finally, the most recent and exhaustive analyses by Miller and Boulter
\cite{Boulter}, Grassberger \cite{QGrassberger}, and Drossel and
coauthors \cite{WD} conclude unambiguously, using a variety of
arguments and large-scale computer simulations, that {\it the
  spatially extended version of the non-conserving OFC model is not
  critical}.

As a consequence, the state-of-the-art is that, despite of the apparent power
law distributions spanning for a few decades, {\it the OFC model is not truly
  scale-invariant, except for its conservative limit}.  The question of
whether real earthquakes are described or not by this type of SOC models or
other type of mechanisms need to be invoked remains unsolved
\cite{Yang,Her,Ramos}.

\subsubsection{Forest fires:}

The Drossel-Schwabl model \cite{FFM2} (see also \cite{FFM2_Henley}) is an
improved version of an older forest-fire model \cite{oldFFM} proposed to
explain the apparent scale-invariance of real forest-fires \cite{realfires}
(see Appendix A).

Three type of states are defined: $z=0$ or empty, $z=1$ or occupied by a tree,
and $z=2$ or burning tree. At each time step, new trees grow, at rate $p$, at
randomly chosen sites provided they were empty, and trees catch fire at a much
smaller rate $f$.  Fire propagates deterministically to neighboring occupied
sites and, after burning, trees become empty sites. The relevant parameter is
$1/\theta=p/f$ \cite{theta}, and the model has been claimed to be critical
provided that the double limit $f \rightarrow 0$, $p \rightarrow 0$ with $f/p
\rightarrow 0$) is taken \cite{FFM2}.

Observe that {\it a double separation of time scales} is imposed in the model
definition: trees are born at a much faster rhythm than fires occur and fires
propagate at a much faster pace than trees grow \cite{FFM2}.  This is to be
compared with the single time scale separation in sandpiles \cite{FFM2,GG}. We
shall discuss later the consequences of such a double separation of
timescales.

Analytical results and mappings into a branching process \cite{Harris} first
suggested some similarities with standard percolation models
\cite{FFM2,FFM2_MF,FFM_Grassberger}. Given the limited analytical tractability
of these models, the controversy about the existence of true criticality was
mainly played on the ground of computer simulations
\cite{FFM2,FFM2_Romu,FFM_Grassberger,FFM2_Grassberger}.  For sufficiently
large systems, anomalies were reported to appear; among them: {\it (i)} the
repulsive character of the fixed point ($f/p=0$), {\it (ii)} the coexistence
of largely-subcritical and supercritical clusters of trees, {\it (iii)} the
existence of two length scales with different exponents into the system, {\it
  (iv)} the violation of standard scaling for the distribution of avalanche
sizes $P(s)$, and {\it (v)} a pathological finite-size behavior
\cite{FFM2,FFM_Grassberger,FFM2_Romu,FFM2_Grassberger,FFM2_8}.

Finally, when ``massive'' simulations of extremely large systems very close to
the critical regime were accessible \cite{FFM2_Jensen,Mega_Grassberger}, these
anomalies turned into a lack of true critical behavior, beside of the apparent
scaling observed for a few decades: {\it the self-organized stationary state
  of the Drossel-Schwabl model is not critical}.

In the rest of the paper, we shall rationalize and generalize the
above conclusion (i.e. absence of {\it bona-fide} criticality of
earthquake and forest-fire SOC models) to generic non-conserving
systems. to this end, we shall employ three different unified
frameworks, as described in the three forthcoming sections.

\section{A simple mean-field approach}
\label{zeroth}

As already mentioned, the random-neighbor version of the OFC model has been
solved analytically, with the conclusion that, except for the conserving
limit, it is not critical, but generically subcritical \cite{Broker,Hakim}. On
the other hand, in a subsequent work, Pruessner and Jensen \cite{PJ}
considered a modified version of such a model in which, by including a
different (stronger) loading mechanism, they showed that criticality can be
restored in the infinite size limit. In this section, we review the results in
\cite{PJ} and study the finite-size scaling of this and related mean-field
systems.

The model in \cite{PJ} is somewhere in between forest-fire and earthquake
models. It is defined as follows: consider a set of $N$ sites, each of them
with an associated energy $z_{i}$ (with $z$ a continuous variable). As in the
OFC model, three types of states exist: {\it stable}, with an energy $0\leq
z_{i}<1-\alpha$; {\it susceptible}, with $1-\alpha\leq z_{i}<1$; and {\it
  active} sites, with $z_{i} \geq 1$.  The main difference with respect to the
OFC model is that, between avalanches, driving and loading operate as
independent mechanisms \cite{PJ}:
\begin{itemize}
\item {\it Background loading}: in the spirit of forest-fire models
  \cite{FFM2,FFM2_Grassberger}, $(1/\theta)$ sites are randomly chosen and
  their respective energies are increased to $1-\alpha$ (i.e. become
  susceptible) provided they were stable; otherwise, nothing happens.

\item {\it Triggering of an avalanche}: A randomly chosen site, $i$, is
  activated ($z_{i}=1$) provided it was susceptible.
\end{itemize}

The relaxation dynamics within avalanches is identical to that of the
random neighbor OFC model with $m$ random neighbors: sites above
threshold are emptied, $z=0$, and the energy of its (randomly chosen)
neighbors is increased by a fixed amount $\alpha$.  Conservation holds
for $\alpha=1/m$.

At a mean-field level, the condition for stationarity is given by the
following energy-balance equation \cite{PJ}:
\begin{equation}
(1-m\alpha)z_{a}\langle s\rangle=(1-z_{s})+
(1/\theta)(1-\alpha-z_{st})\dfrac{\zeta_{st}}{\zeta_{s}},
\label{balance}
\end{equation}
where the different terms are as follows:

({\it i}) l.h.s: For each relaxation event at site $i$, the amount of
dissipated energy is $(1-m \alpha)z_{i}$; hence, the average
dissipation during an avalanche is $(1-m\alpha)z_{a}\langle s\rangle$,
where $ \langle s \rangle$ is the average avalanche size and $z_{a}$
the average energy of active sites.

({\it ii}) r.h.s. first term: Triggering increases the energy of the selected
susceptible site, $i$, by an amount $1-z_{i}$.  The corresponding average
increase is $1-z_{s}$, where $z_{s}$ is the average energy of susceptible
sites.

({\it iii}) r.h.s., second term: Every time the background is loaded, the
energy of a stable site, $i$, is increased by an amount $(1-\alpha-z_{i})$.
This is attempted $(1/\theta)$ times and, the average number of triggering
events before an avalanche is actually generated is $\zeta_{st}/\zeta_{s}$
(where $\zeta_{st}$ and $\zeta_{s}$ are the density of stable and susceptible
sites, respectively).  The average increase of background energy per avalanche
is, finally, $(1/\theta)(1-\alpha-z_{st})(\zeta_{st}/\zeta_{s})$, where
$z_{st}$ stands for the average energy of stable sites.

In this way, Eq.(\ref{balance}) establishes that, for a steady state
to exist, the average dissipated energy should be compensated by the
averaged energy increase of driving and loading.  Now, imposing in
Eq.(\ref{balance}) that $(1/\theta)$ diverges, one of the following
two conditions must be obeyed for Eq.(\ref{balance}) to hold
\cite{PJ}:
\begin{enumerate}
\item $ (1-m\alpha)=0$, i.e.  there is strict conservation, or
\item $\langle s \rangle$ diverges, which is a necessary condition for
  criticality.
\end{enumerate}
In the second case, by studying the probability distribution function of $z$
and using a mapping into a branching-process, it has been shown that not only
$\langle s \rangle$ diverges, but also that the system is {\it critical in the
  infinite size limit} \cite{PJ}. Moreover, as expected for a mean-field
model, the size-avalanche exponent is found to be $\tau=3/2$
\cite{Bak,Jensen,Moloney,MF}.

However, as already pointed out in \cite{PJ}, for any finite system size, $N$,
neither $1/\theta$ nor $\langle s \rangle$ are infinite. In such a case, a
finite value of the parameter $1/\theta$ {\it must be fine-tuned} to some
precise value, $1/\theta_c(N)$, for Eq.(\ref{balance}) to hold.  Such a value
should diverge slower than $N$, in order to achieve the right limit
$(1/\theta_c(N))/N \rightarrow 0$ for $N \rightarrow \infty$, but there is no
analytical prescription in \cite{PJ} on how to fix it for each system size.

This is a well-known problem, shared by forest-fire models, where the number
of trees grown between two fires ($1/\theta$) is a parameter which needs to be
carefully tuned for any finite size: too small values lead to subcritical
fires, while too large values generate super-critical fires spanning the whole
system (and generating a bump for large values in the size distribution).

\begin{figure}
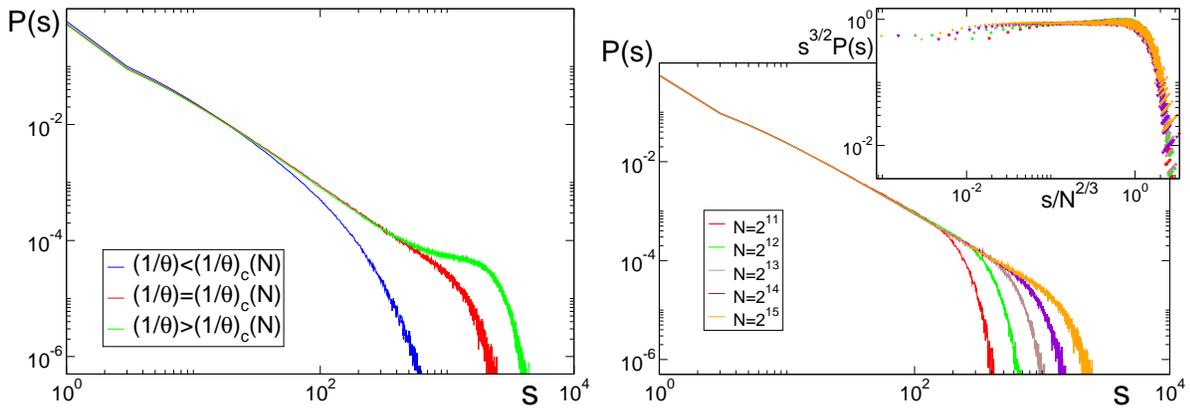

\begin{center}
  \includegraphics[height=5.3cm]{PJ1.eps}
\includegraphics[height=5.3cm]{PJ2.eps}
\caption{Left: Avalanche size distribution for the random-neighbor model
  introduced by Pruessner and Jensen \cite{PJ}, for $N= 2^{15}$ and
  dissipation parameter $\alpha= 0.15$. Subcritical, almost-critical (but
  still cut-off by system-size) and supercritical distributions, corresponding
  to different values of the loading parameter ($1/\theta=2$, $8$, and $22$,
  respectively) are shown. Right: Avalanche size distribution at the {\it
    fine-tuned} critical values of $1/\theta_c(N)$ for different sizes (from
  $2^{11}$ to $2^{15}$); the corresponding cut-offs grow with system size in a
  scale-invariant way. In the inset, the same data are collapsed into a unique
  scaling curve by using Eq.(\ref{scaling2}).}
\label{PJ}
\end{center}
\end{figure}

This is graphically illustrated in Fig.~\ref{PJ} (left), where the avalanche
size distribution for the Pruessner-Jensen model is plotted for a system with
$ N=2^{15}$ sites, $\alpha=0.15$, and three different values of $1/\theta(N)$;
even if the values of $1/\theta(N)$ are large, the size-distributions are not
pure power-laws: they are either subcritical (with an exponential cut-off) or
supercritical (with a bump for large avalanches).  Nevertheless, partial
scaling is observed in any case.
Given that the control parameter $1/\theta$ is an integer number, criticality
cannot be tuned with arbitrary precision (specially for small system sizes)
but, still, for each value of $N$ it is possible to find an almost critical
value of $1/\theta$, $1/\theta_c(N)$. In Fig.~\ref{PJ} (right) we show the
avalanche size distribution at the fine-tuned critical point for different
values of $N$.  Increasing the system size we have observed that such a value
scales as $ 1/\theta_c(N) \sim N^{0.33(5)}$ suggesting
\begin{equation}
  1/\theta_c(N) \sim  N^{1/3}.
\label{scaling}
\end{equation}
A collapse of the critical size-distribution curves for different values of
$N$ (see the inset of Fig.~\ref{PJ}) leads to $ P(s,N) \sim s^{-1.5(1)}
\exp(-s/N^{0.65(5)})$, compatible with
\begin{equation}
  P(s,N) \sim  s^{-3/2} \exp(-s/N^{2/3}),
\label{scaling2}
\end{equation}
where the mean-field exponent $ \tau=3/2$ is recovered.

In Section \ref{Langevin}, we shall introduce a scaling Langevin theory for
non-conserving SOC models, which explains in a straightforward way the two
(formerly unknown) scaling laws, Eq.(\ref{scaling}) and Eq.(\ref{scaling2}).

Let us comment on the peculiarity of the thermodynamic limit in this model (as
well as in the Drossel-Schwabl forest-fire): the condition $1/\theta_c(N)
\rightarrow \infty$ is automatically fulfilled by imposing the double
separation of time-scale discussed above, {\it i.e.} the second separation of
timescales is tantamount to fine-tuning $\theta_c({\infty})$ to its critical
value $\theta_c(\infty)=0$.  Note also that the infinite size limit is somehow
pathological as the entire super-critical phase (as well as the critical point
itself) collapses into a unique single point $\theta_c(\infty)=0$.  Instead,
for finite systems, a precise (not infinite) double separation of scales is
required to set the system to the critical point, separating distinct
sub-critical and super-critical phases. This boils down to the need of
fine-tuning $1/\theta_c(N)$ for each value of $N$ to have a coherent
finite-size scaling.

On the contrary, in the conserving limit, large avalanches spanning the whole
system are observed for any size and criticality is reached without resorting
to careful tuning.

Summing up, even if the random-neighbor model studied in \cite{PJ} exhibits
infinite avalanches and is critical in the infinite system size limit, it
lacks of a well-defined {\it finite size scaling} and, therefore, it is not
truly scale-invariant: for any finite system, deviations from criticality are
observed if the control parameter $1/\theta(N)$ is not fine tuned to a precise
$N$-dependent critical value. In conclusion, this model does not qualify as a
{\it bona fide} self-organized critical system.

In this respect, the situation for the RN-OFC model studied in
\cite{Broker,Hakim} is even worse: given that it lacks a parameter analogous
to $1/\theta$ to be tuned, the degree of loading cannot be regulated and
the model is generically subcritical for any non-vanishing dissipation rate
even in the large system size limit \cite{details}.

\section{Self-organized mean-field approach: Self-Organized
  Branching Process}
\label{SOBP}

In this section, we complement the mean-field approach of the previous
one by exploiting the {\it self-organized branching process}
introduced by Zapperi, Lauritsen, and Stanley in \cite{SOBP}.  First,
in Subsection \ref{1}, we introduce this approach for a conserving
sandpile model (the Manna cellular automaton). Then, following also
Zapperi {\it et al.}, in Subsect.  \ref{2} we move on to analyze
dissipative models, showing that they are generically subcritical.
Finally, as a last step, in Subsect. \ref{3} we implement a
loading mechanism in the self-organized branching process which
captures the essence of earthquake and forest-fire models, and explore
under which circumstances the resulting model is critical.

\subsection{Conserving case}
\label{1}
In sufficiently high spatial dimensions (i.e. in the mean-field regime),
avalanches in sandpiles rarely visit twice the same site; activity patterns
are mostly tree-like. An avalanche can be seen as a {\bf branching process}
\cite{Harris} in which an individual ({\it ancestor}) creates a fixed number
$k$ of {\it descendants} with probability $p(k)$. The average number of
descendants per ancestor, $\sigma=\sum_{k}kp(k)$, is called {\bf branching
  ratio}. For $\sigma>1$ avalanches propagate indefinitely (super-critical
phase), for $\sigma<1$ they stop after a typical number of generations
(sub-critical phase), while the process is critical in the marginal case,
$\sigma_c=1$ \cite{Harris}.

\begin{figure}
\begin{center}
  \includegraphics[height=6.0cm]{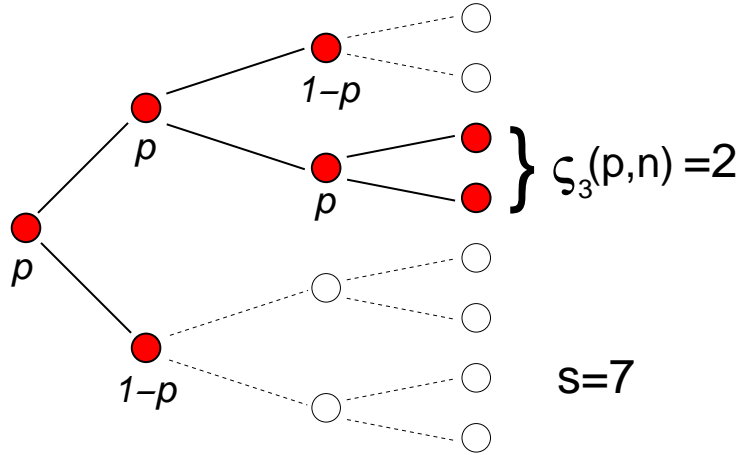}
  \caption{\footnotesize{Left: Sketch of the propagation of an avalanche in a
      self-organized branching process of size $N=2^{3+1}-1=15$. The avalanche
      has size $s=7$ (red spots), and has reached the border (i.e. there are
      two occupied sites in the last generation; $\varsigma(p,n)=2$). Figure
      adapted from \cite{SOBP}}}
\label{sketch_SOBP}
\end{center}
\end{figure}

In this {\it static branching process}, fixing, without loss of generality,
the number of descendants to $k=2$, a given active site branches in $2$ with
probability $p$ or has no offspring with probability $1-p$ (see
Fig.~\ref{sketch_SOBP}), ensuing $\sigma=2p$ and a critical value
$p_{c}=1/2$.

Let us consider, as a simple example, the Manna sandpile with critical
threshold $z_{thr}=2$ (see Appendix A), and map its mean-field version into a
{\it self-organized branching process}, in which $p$ itself is a dynamical
variable \cite{SOBP}. In the Manna dynamics, each grain arriving at site $i$
can either generate activity (energy above threshold) if previously $z_{i}=1$,
or not if $z_{i}=0$.  An avalanche in high spatial dimensions can be,
therefore, seen as a branching process with $p=P(z=1)$, i.e. the background
energy density is nothing but the branching ratio \cite{SOBPCO}.  A {\it
  generation} is defined as the set of sites probed for activation at each
time-step; after $m$ generations there are $N=2^{m+1}-1$ involved sites (see
Fig.~\ref{sketch_SOBP}).
For computer simulations, we fix a maximum number of generations $m$ and
impose (to mimic boundary dissipation) that at the $m$-th generation all
grains are lost. Note that, apart from such a dissipative boundary, the bulk
dynamics is conserving.

\begin{figure}
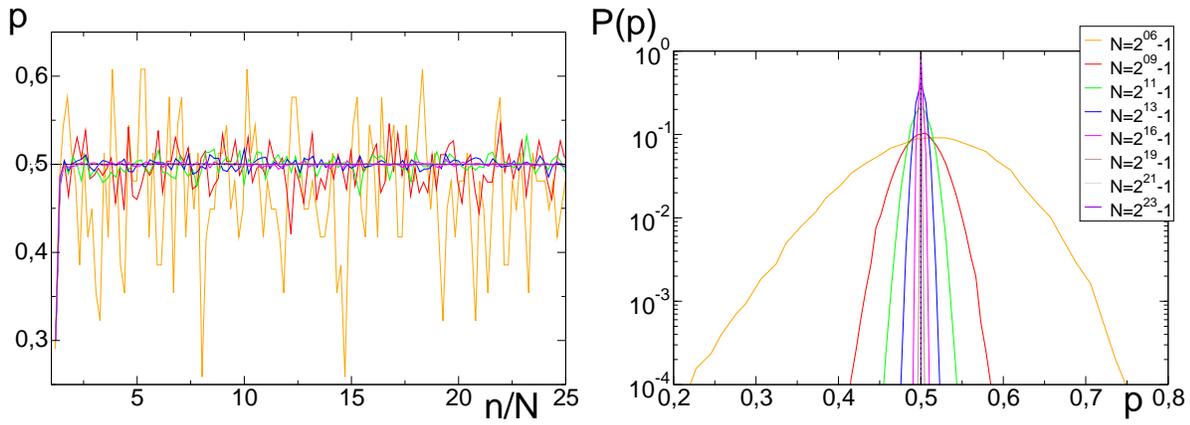

\begin{center}
\includegraphics[height=5.5cm]{SOBP2.eps}
\includegraphics[height=5.5cm]{SOBP3.eps}
\caption{\footnotesize Left: Evolution of the branching probability $p(n)$ in
  a computer simulation of the self-organized branching process as a function
  of the number of avalanches per site, for different sizes $N$ (ranging from
  $2^6-1$ to $2^{23}-1$). In all cases, $p$ fluctuates around $p^{*}=1/2$; the
  larger the system, the smaller the amplitude of fluctuations.  Right:
  Stationary probability distribution functions of values of $p$ for various
  sizes (same color code in both figures); the width decreases with increasing
  system-size and vanishes in the thermodynamic limit.}
\label{SOBP-evol}
\end{center}
\end{figure}

Each avalanche modifies the background in which the next avalanche is to be
started; i.e. it changes the value of $P(z=1)$. Thus, the branching
probability $p$ becomes a fluctuating variable, as illustrated in
Fig.~\ref{SOBP-evol}. In the left part of the figure, the value of $p$ is plot
as a function of the avalanche number for different system sizes, while in the
right figure the statistically stationary distribution of values of $p$ is
represented for various sizes; the width of the distributions decreases with
increasing size and can be made as small as wanted.

To recover analytically these computational observations, let $Z(n)$ be the
total number of grains into the system after $n$ avalanches; then,
$p(n)=Z(n)/N$. If $\varsigma(p,n)$ is the number of grains dissipated at the
$m$-th (last) generation of the $n$-th avalanche, in order to have
stationarity, the following balance equation:
\begin{equation}
  Z(n+1)=Z(n)+1-\varsigma(p,n),
\label{SOBP-balance}
\end{equation}
or
\begin{equation}
p(n+1)=p(n)+\dfrac{1-\varsigma(p,n)}{N},
\label{SOBP_p_eq}
\end{equation}
must hold.  The average number of grains dissipated at the boundary is
$\langle \varsigma(p,n) \rangle=(2p)^{m}$ ($2^m$ sites at the
boundary, each one occupied with probability $p$ \cite{Harris}).  For
each avalanche, $\varsigma(p,n)=(2p)^{m}+\eta(p,n)$, where $\eta(p,n)$
is a Gaussian white noise. Plugging this into Eq.(\ref{SOBP_p_eq}) and
taking the continuum limit for $n$, one can formally write:
\begin{equation}
  \dfrac{dp}{dn}=\dfrac{1-(2p)^{m}}{N}+\dfrac{\eta(p,n)}{N},
\label{SOBP_p_eq_def}
\end{equation}
whose deterministic part has a stable fixed point at $p^{*}=1/2=p_{c}$.
Accordingly, in the thermodynamic limit (in which the effect of fluctuations
can be neglected \cite{SOBP}), the dynamics attracts $p$ to its critical value
(see Fig.~\ref{SOBP-evol} left) and the width of the fluctuations of $p$
around $p=p^*=p_c$ decreases with increasing system size (see
Fig.~\ref{SOBP-evol} right). This simple (conserving) branching process
self-organizes to its critical point.

\subsection{Dissipative case}
\label{2}
Borrowing still from Lauritsen, Zapperi and Stanley \cite{SOBP2}, let us
introduce a non-vanishing bulk-dissipation rate into the Manna model and, as a
consequence, into its self-organized branching process representation. Each
offspring (not necessarily in the last generation) is removed from the system
with probability $\epsilon$; the effective branching probability becomes
$q=p(1-\epsilon)$, and the criticality condition is $q=1/2$, or
\begin{equation}
p_{c}=\dfrac{1}{2(1-\epsilon)},
\label{new_pc}
\end{equation}
 and $ \langle \varsigma(q,n)
 \rangle=(2q)^{m}=\left(2p(1-\epsilon)\right)^{m}$.  Eq.(\ref{SOBP_p_eq})
 transforms into:
\begin{equation}
  p(n+1)=p(n)+\dfrac{1-\varsigma(q,n)-\kappa(q,n)}{N}.
\label{SOBPD_p_evol}
\end{equation}
where $\kappa(q,n)$ is the total amount of grains dissipated in the bulk.  A
simple calculation, synthesized in Appendix B, leads to the following equation
for the evolution of $p$
\begin{equation}
  \dfrac{dp}{dn}=A(p;\epsilon;N)+\dfrac{\eta(p,n)}{N},
\label{Juanico_p_SOBP}
\end{equation}
where
\begin{equation}
%  \begin{array}{rl}
 A(p;\epsilon;N) =
%&
\dfrac{1-\left(2p(1-\epsilon)\right)^{m}}{N} 
%\\
%    &\\
-
%&
\dfrac{p\epsilon}{N\left[(1-p)+p\epsilon\right]}
\left[1+\dfrac{1-\left(2p(1-\epsilon)\right)^{m+1}}{1-2p(1-\epsilon)}
  -2\left(2p(1-\epsilon)\right)^{m}\right]
%\\ &\\ + & \dfrac{\eta(p,n)}{N}, \end{array}
\label{Long}
\end{equation}
and, as above, the noise amplitude is N-dependent. After some simple algebra
and omitting the noise term, Eq.(\ref{Juanico_p_SOBP}) can be rewritten as
\begin{equation}
\dfrac{dp}{dn}= \frac{1-x^m}{(1-x)~N} (1-2p)
\label{short}
\end{equation}
with $x=2p(1-\epsilon)$. It is straightforward to check that the only stable
fixed point of Eq.(\ref{short}) is $p^{*}=1/2$ \cite{fixed}.  Therefore, as
$p_{c}=1/2(1-\epsilon)$, the self-organized dynamics leads to a {\it
  sub-critical} point $p^* = ~1/2 < p_c$; the fixed-point branching ratio is
less than unity, and the process propagates only for a finite number of
generations for any non-vanishing value of $\epsilon$ \cite{SOBP2}. It is only
in the conserving limit, $\epsilon=0$, that the self-organized value and the
critical point coincide; otherwise there is {\it self-organization to a
  sub-critical point}.

\subsection{Dissipation and loading}
\label{3}

In a recent series of papers \cite{Juanico}, Juanico and collaborators
introduced a background dynamics into the self-organized branching model with
dissipation. These authors consider a dissipative version of the Manna
sandpile rules in the following way: with probability $\alpha$, an active site
transfers $2$ grains to $2$ different randomly chosen neighbors; with
probability $\beta$, only one grain is transferred to one neighbor while the
other one is dissipated; finally, with probability $\epsilon=1-\alpha-\beta$,
the two toppling grains are dissipated. For simplicity and to easy comparison
with the calculations above, we fix $\beta=0$; the critical branching
probability becomes $p_c=1/2(1-\epsilon)=1/2\alpha$.

\begin{figure}
\begin{center}
  \includegraphics[height=6.0cm]{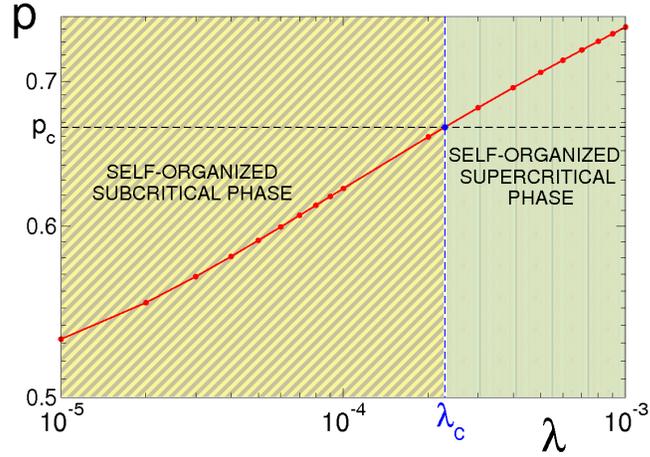}
  \caption{Fixed point of the evolution of Eq.(\ref{flow2}) in the
    self-organized branching process as a function of $\lambda$, for a fixed
    dissipation parameter $\epsilon=0.25$.  Note that only for
    $\lambda=\lambda_c(\epsilon) \approx 2.288 ~ 10^{-4}$, the fixed point
    corresponds to the critical value of $p$,
    $p_c(\epsilon)=1/2(1-\epsilon)=2/3$. Otherwise, the system self-organizes
    to a subcritical or to a supercritical point.}
\label{Juanico1}
\end{center}
\end{figure}

A {\it background dynamics} is implemented in \cite{Juanico} by introducing a
rate $\lambda$ for a stable site to be turned into a critical one ($z=0
\rightarrow z=1$), and a rate $\vartheta$ for the opposite
transformation. From now on, and without loss of generality, we restrict
ourselves to the ``loading'' process (which increases the energy) and fix
$\vartheta=0$. Neglecting the noise term, it is straightforward to arrive at
the following evolution equation for $p$,
\begin{equation}
\dfrac{dp}{dn}=(1-p) \lambda + \frac{1-x^m}{(1-x)N} (1-2p),
\label{flow2}
\end{equation}
with $x=2p\alpha$. It has a stable fixed point at
\begin{equation}
\lambda = \frac{x^m-1}{(1-x)N} \frac{1-2p}{1-p}.
\label{tuning}
\end{equation}
Note that, at $p_c$ $x$ is equal to $1$ and thus, if $\lambda$ is {\it
  fine-tuned} to
\begin{equation}
\lambda_c = \frac{m}{N} \frac{2p_c-1}{1-p_c},
\label{lambda}
\end{equation}
then the fixed point of Eq.(\ref{flow2}) becomes $p_c$, i.e. the pair
$(\lambda_c,p_c)$, with $\lambda_c$ given by Eq.(\ref{lambda}) and
$p_c=1/2\alpha$, fulfills Eq.(\ref{tuning}) and the self-organized branching
process becomes critical. On the other hand, fixing $\lambda > \lambda_c$
(resp.  $\lambda < \lambda_c$) the fixed point becomes supercritical
(resp. subcritical) as illustrated in Fig.~\ref{Juanico1}.  In conclusion, by
{\it carefully tuning} the loading parameter to exactly compensate the
effect of dissipation, the system self-organizes to its critical point. This
process, requiring an explicit parameter tuning, {\it cannot be called bona
  fide self-organization}.
Finally, note that both, the critical loading parameter $\lambda_c$ and
the driving rate $1/N$, vanish in the large system size limit ($m \rightarrow
\infty, N \rightarrow \infty$), while the ratio ``loading over driving''
($\lambda_c = m\frac{2p_c-1}{1-p_c}$) diverges. These conditions are analogous
to those usually imposed to forest-fire models ($p, f \rightarrow 0, p/f
\rightarrow \infty$). The calculation above illustrates that such conditions
are necessary but not sufficient to achieve criticality in the absence of a
conservation law in any finite system: a size-dependent fine tuning is also
required.

Even if Juanico and coauthors claim to have designed critical self-organized
models, all branching processes studied by them \cite{Juanico} are similar in
spirit to the example above: close inspection of their rules reveals an
underlying parameter fine tuning in all the different variations they
study. For example, for the full model described above (i.e.  with $\beta \neq
\vartheta \neq 0$), Juanico et al.  explicitly make the ``convenient'' choice
of parameters
\begin{equation}
\dfrac{\vartheta}{\lambda}=2\alpha+\beta-1=\dfrac{1}{p_{c}}-1,
\label{Juanico_tuning}
\end{equation}
analogous to Eq.(\ref{lambda}). Actually, in \cite{Juanico},
Eq.(\ref{Juanico_tuning}) is obtained by fine-tuning the
$N$-independent terms to the critical point, while the last term in
the flow equation, proportional to $1/N$, is neglected. Not
surprisingly, the resulting model converges to its critical point in
the thermodynamical limit; but, requiring careful tuning, it cannot be
properly called critical self-organization.

{Summing up}, in this section we have illustrated that the conserving
self-organized branching process shows asymptotically critical dynamics, while
its dissipative counterpart self-organizes to a subcritical point. Introducing
a loading mechanism, dissipation can be compensated and criticality restored
{\it if and only if} the loading rate is fine-tuned to a precise
size-dependent value.  Otherwise, the system self-organizes generically either
to a subcritical point or to supercritical one.

%%%%%%%%%%%%%%%%%%%%%%%%%%%%%%%%%%%%%%%%%%%%%%%%%%%%%%%%%%%%%%%%%%%%%%%
\section{A full description: Langevin theory of SOC}
\label{Langevin}

Having already studied two different mean-field like approaches, in
this section we discuss the complete Langevin theory of
self-organizing systems.  This theory explains the origin of the
underlying critical point beyond mean-field, its universality in any
dimension, as well as the key mechanism producing SOC.

First (Subsection \ref{51}), we review the existing
absorbing-phase-transition Langevin picture of conserving SOC models.
Then (Subsect. \ref{52}), we introduce bulk-dissipation and extend the
theory to non-conserving systems. Finally (Subsect.  \ref{52}), a
loading mechanism is introduced to elucidate the behavior of
non-conserving self-organized systems. We emphasize the substantial
differences with respect to the conserving case.

\subsection{Langevin theory with conservation \label{CDP}}
\label{51}

The main idea to construct a stochastic theory of conserving SOC is to
``regularize'' sandpiles (and related systems) by switching off both boundary
dissipation and slow driving \cite{BJP,MF,FES0,FES,EarlyFES}.  In this way,
the total amount of sand or ``energy'', $E$, in the pile becomes a conserved
quantity, and can be retained as a control parameter. Indeed, in the
so-defined ``{\it fixed-energy ensemble}'' and for large values of $E$, the
system is in an {\it active phase} with never-ending relaxation
events. Instead, for small values of $E$ it gets trapped with certainty into
some {\it absorbing state} \cite{AS} where all dynamics ceases (i.e. all sites
are below threshold).  Separating these two regimes there is a critical
energy, $E_c$, at which an absorbing phase transition takes place.  In this
way self-organized criticality is related to a standard phase transition
\cite{BJP,Ivan,Broker2}.

It has been shown \cite{BJP,FES,Chris04} that such a critical value,
$E_c$, coincides with the stationary energy density to which the
original self-organizing sandpile converges. In other words, the
energy around which the standard sandpile (i.e. including slow-driving
and boundary dissipation) fluctuates is the critical point of the
``fixed-energy sandpile''. Furthermore, the width of fluctuations
becomes smaller and smaller with increasing system size (see left
part of Fig.~\ref{Panels}), guaranteeing that in the thermodynamic
limit the original sandpile self-organizes to criticality.

This connection between ``driven/dissipative'' systems and their
``fixed-energy'' counterparts permits us to relate avalanche exponents to
standard critical exponents (see \cite{avalanches} for scaling relations) and
to {\it rationalize the critical properties of SOC systems from the broader
  point of view of standard non-equilibrium (absorbing-state) phase
  transition} \cite{BJP,FES,debate}.

Using this approach, it has been established that stochastic sandpiles {\it do
  not} belong to the robust {\it directed percolation (DP) class}, prominent
among absorbing phase transitions, but to the so-called ``conserving-DP''
({\it C-DP} hereafter) or Manna class.  This class is characterized by the
coupling of activity to a static conserved field representing the conservation
of sandgrains \cite{BJP,FES,Romu,Lubeck}.  The field theory or set of
mesoscopic Langevin equations proposed under phenomenological grounds to
describe this class is:
\begin{equation}
\left\{
\begin{array}{lll}
  \partial_t \rho(\vec{x},t) &=& a\rho-b\rho^{2}+\omega\rho E
  +D\nabla^2\rho+\sigma\sqrt{\rho}\eta ({\vec{x}},t)\\ 
\partial_{t} E({\vec{x}},t) &=&
  D_{E}\nabla^2\rho({\vec{x}},t)
\end{array}
\right.
\label{FES}
\end{equation}
where $\rho({\vec{x}},t)$ is the activity field (characterizing the
density of grains above threshold), $E({\vec{x}},t)$ is the
locally-conserved energy field, $a, b, \omega, D, \sigma$, and $D_{E}$
are parameters and $\eta (\vec{x},t)$ is a Gaussian white noise.  Some
dependences on $({\vec{x}},t)$ have been omitted to unburden the
notation.

Note that in the C-DP class, two fields are required for a Langevin
representation: the activity field representing grains/energy/force {\it above
  threshold} and the background or energy field describing the local amount of
grains/energy/force.  Nevertheless, it is possible to stick to a single field
description by integrating out the energy equation. This generates two extra
terms for the activity equation
\begin{equation}
  \omega \rho E(\vec{x}, 0) - \omega \rho \int_0^t dt' \nabla^2 \rho(t').
\label{nonmarkovian}
\end{equation}
The second, history dependent (non-Markovian) term describes the tendency of
sites that have been less active than their neighbors in the past to be more
susceptible for activation (e.g. in a sandpile, if the neighbors of a given
site have toppled, the site is very likely to overcome the threshold). The
single equation for the activity reads
\begin{equation}
\begin{array}{lll}
  \partial_t \rho(\vec{x},t) &=& (a + \omega E(\vec{x}, 0))\rho-b\rho^{2}-
  \omega \rho \int_0^t dt' \nabla^2 \rho({\vec{x}},t')
  +D\nabla^2\rho+\sigma\sqrt{\rho}\eta ({\vec{x}},t);
\end{array}
\label{FES2}
\end{equation}
non-Markovianity is the price to pay for removing the energy field
\cite{quenched}.

The C-DP class described either by Eq.(\ref{FES}) or by Eq.(\ref{FES2}) has a
critical dimension $d_c=4$ and embraces not only stochastic sandpiles, but
also (among other examples) some conserving reaction-diffusion systems, for
which the equations above can be explicitly derived from the microscopic
dynamics \cite{Romu,Lubeck}.

Eq.(\ref{FES}) can be studied either ({\it i}) in spatially extended systems,
({\it ii}) using random neighbors, or ({\it iii}) in a globally, all-to-all,
coupled version in which the Laplacian is replaced by $
\bar{\rho}-\rho(\vec{x},t) \simeq \frac{1}{N}\sum_{y \neq x} [\rho(\vec{y},t)
  - \rho(\vec{x},t)]$.  These last two are useful to construct mean-field
approximations of the full (spatially extended) theory.

\subsubsection{Relation with other universality classes}

First, note that the well-known theory for {\it directed percolation}
(i.e. the Reggeon field theory \cite{AS}) is recovered upon fixing $\omega=0$
in Eq.(\ref{FES}).  The additional conserved field turns out to be a relevant
perturbation altering the critical behavior of systems in the DP class
\cite{BJP,FES,Lubeck}.

On the other hand, for the sake of completeness, let us just briefly mention
that the C-DP class is fully equivalent to the pinning/depinning transition of
interfaces in random media, i.e. the {\it Quenched Edwards Wilkinson}
\cite{Mikko,Interfaces,Jabo}. The absorbing (resp. active) phase maps into the
pinned (resp. depinned) one.  Exploiting the mapping between these two
descriptions (of a unique underlying physics), critical exponents for
Eq.(\ref{FES}) can be deduced from existing renormalization group results for
interfaces \cite{LIM-renorm}.

It is worth stressing that, in terms of interfacial models, {\it
  conservation of energy is equivalent to interface translation
  invariance}; for instance, a dissipative term like $-\epsilon \rho$
introduced in Eq.(\ref{FES}) would map into a term $-\epsilon h$
(where $h=h(\vec{x},t)$ is the interface height) in the Quenched
Edwards Wilkinson equation, which breaks such an invariance.

In this respect, recent experimental evidence of self-organized critical
behavior (including finite size scaling), obtained for {\it avalanches in type
  II superconductors} \cite{superc}, give critical exponents compatible with
those of the C-DP class.  Barkhausen noise \cite{Barkhausen} and acoustic
emission in fracture \cite{acoustic} are other related examples.

\subsubsection{Conserved SOC as an absorbing state transition.}

\begin{figure}
\begin{center}
 \includegraphics[height=5.5cm]{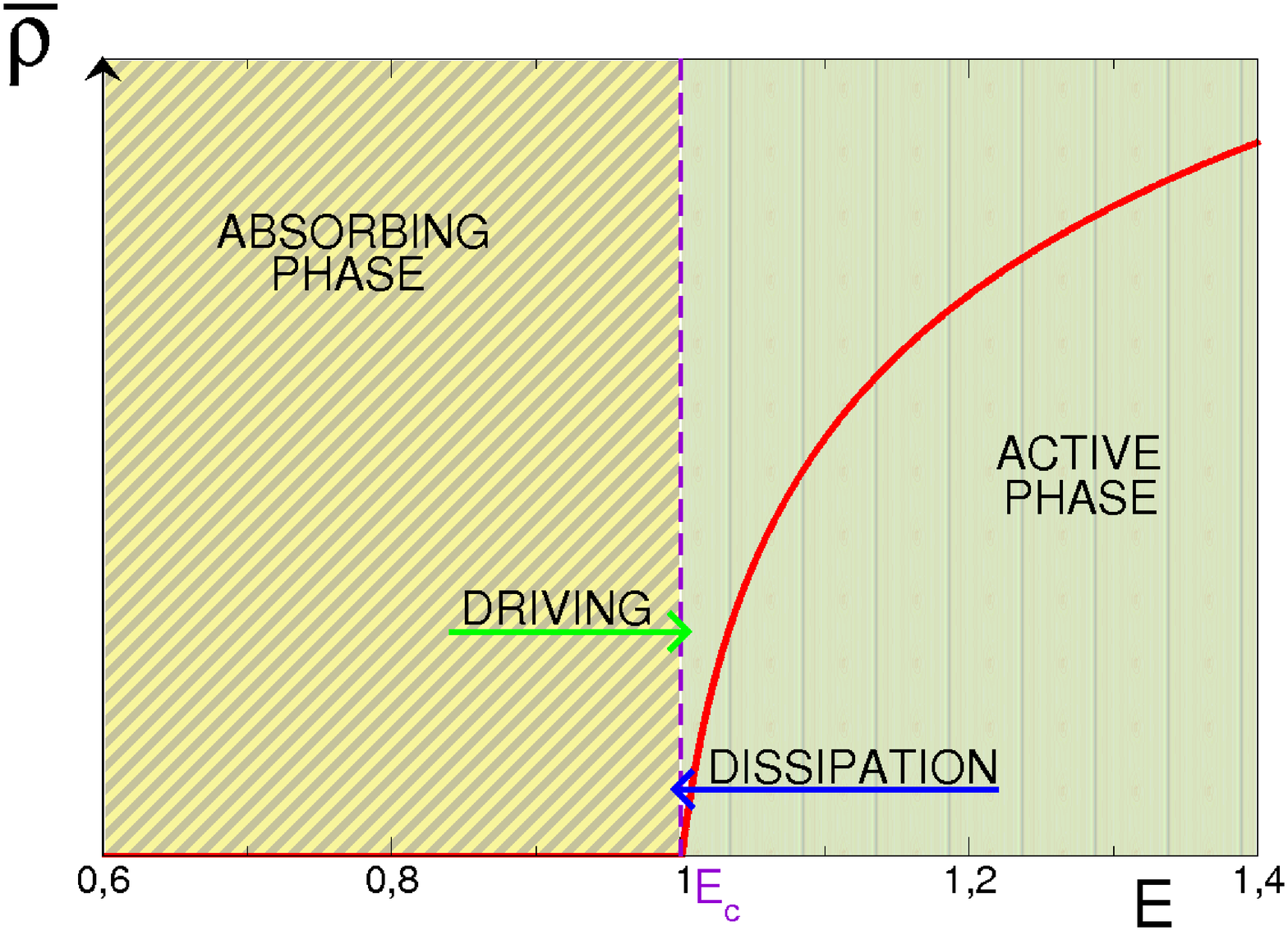}
\includegraphics[height=5.5cm]{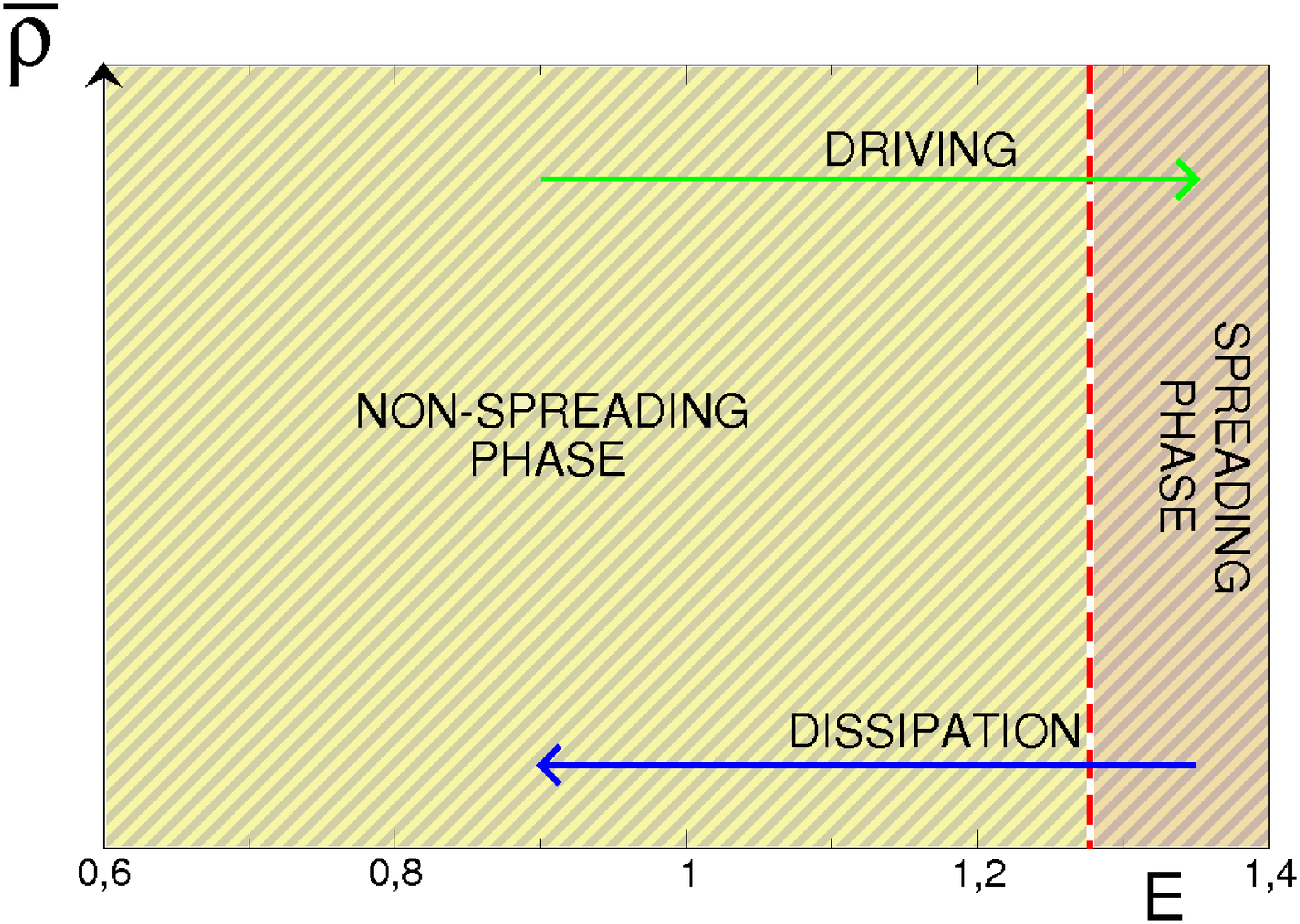}
\\
\includegraphics[height=5.4cm]{C3.eps}
\includegraphics[height=5.4cm]{C4.eps}
\caption{(Upper part) Sketched diagrams illustrating the mechanism of
  self-organization for conserving dynamics (left panels) and for
  non-conserving dynamics with loading (right panels).  {\it Left-up}: In the
  ``fixed-energy'' ensemble, there is an active and an absorbing phase,
  separated by a critical point, $E_c$. Slow driving and dissipation make
  sandpiles (and related conserving systems) fluctuate around their associated
  fixed-energy counterpart critical point.  {\it Left-down}: the distribution
  of $E$ during avalanches is plot for simulations of the conserving Langevin
  theory, Eq.(\ref{FES}), using various system sizes ($2^{11}, 2^{12}$, and
  $2^{13}$).  Other parameter values are $a=0.423$, $ b=\omega=1$,
  $D=D_2=0.25$, $ \sigma=\sqrt{2}$; the distributions become progressively
  peaked around $E_c$ upon increasing system size (the value of $E$ has been
  normalized with $E_c$ for each size).  {\it Right}: Plots analogous to their
  counterparts to the left, but for non-conserving systems.  {\it Right-up}:
  In this case, there is no active phase, but just a phase in which
  perturbations/avalanches spread. Results in {\it right-down} correspond to
  simulations of Eq.(\ref{dFES}), using Eq.(\ref{repob}) and Eq.(\ref{seed}),
  with $E_{max}=1.3$ (rest of parameters as above). Observe the much broader
  distributions appearing in this case: the system hovers around the spreading
  critical point with large excursions into both, the supercritical and the
  subcritical phase.}
\label{Panels}
\end{center}
\end{figure}

Within this framework, the way self-organized criticality works is as follows
(see left part of Fig.~\ref{Panels}) \cite{BJP,FES}: if the sandpile is in its
absorbing phase ($E< E_c$, where $E$ stands for the spatially averaged value
of $E(\vec{x})$ in the steady state) then, owing to the driving mechanism, the
energy is slowly increased until, eventually, the {\it active phase} ($E>
E_c$) is reached. At this point avalanches are triggered and they re-structure
the sandpile energy configuration.  Avalanches may dissipate energy at the
open boundaries, until eventually the system falls back into an {\it absorbing
  state}, the avalanche stops, and slow driving acts again restarting the
cycle.  In this way, the sandpile is expected to fluctuate around its critical
point, $E_c$, with excursions to either the active or the absorbing phase as
sketched in the upper-left part of Fig.~\ref{Panels}.

To have a numerical confirmation of this, Eq.(\ref{FES}) can be interpreted as
in SOC, i.e. one can implement slow driving and dissipation at infinitely
separated time-scales, and integrate the equation using the efficient
algorithm introduced in \cite{Dornic}.  In particular, one considers an
absorbing configuration ($\rho(\vec{x},t=0)=0$) and open boundaries, then add
a small amount of activity/energy, $\delta$, to a given site, $ \vec{x_0}$:
\begin{equation}
  \rho(\vec{x_0},t=0) \rightarrow \delta, ~~~~ E(\vec{x_0},t=0) \rightarrow
  E(\vec{x_0},t=0) + \delta;
\label{seed0}
 \end{equation} 
 this generates an avalanche, which evolves according to
 Eq.(\ref{FES}). Iterating this process, one obtains the distribution of
 values $E$ sampled during avalanches, shown in the lower-left panel of
 Fig.~\ref{Panels}.  Observe that, as dissipation and driving become
 arbitrarily small by increasing system size (actually, they are
 infinitesimally small in the thermodynamic limit), the degree of penetration
 into the active and absorbing phases is arbitrarily small, the distribution
 of $E$ becomes more and more peaked, and the system is arbitrarily close to
 its critical point.  Moreover, the avalanche exponents measured by means of
 numerical integration of Eq.(\ref{FES}) at such a steady state coincide with
 (or can be related to) those obtained by performing standard fixed-energy
 simulations of Eq.(\ref{FES}) at its critical point \cite{Dornic,Jabo}.
%\newpage
Note the obvious analogies between this picture and the self-organized
branching process described above: $E$ is the equivalent of $p$, i.e.  the
self-organized control parameter; the critical point $E_c$ corresponds to the
critical branching probability $p_c$.

The advantage of Eq.(\ref{FES}) as a theory for SOC with respect to the
self-organized branching process is that, while this last is a mean-field
theory explaining qualitatively self-organization but failing to justify
critical exponents in spatially extended systems, Eq.(\ref{FES}) is a full
theory including fluctuations and spatial-dimensionality. It provides accurate
estimates for avalanche exponents in any dimension and opens the door to field
theoretical analyses.  Furthermore, Eq.(\ref{FES}), considered on a
random-neighbor or an all-to-all coupling, constitutes a sound mean-field
description of conserving SOC, equivalent to those in the preceeding sections.

To end up, note, once again, the essential role played by conservation in this
theory. The underlying phase diagram sketched in the left part of
Fig.~\ref{Panels} relies on the averaged energy being a control parameter.  If
there was a non-vanishing bulk dissipation, the energy would change
continuously during avalanche evolution.  In the next sub-section, we shall
explore how this affects the absorbing-phase-transition picture of SOC.

\subsection{Langevin theory with bulk-dissipation}
\label{52}

To tackle the problem of non-conservation within the absorbing-state
Langevin framework, we need to modify Eq.(\ref{FES}) to allow for bulk
dissipation.  Introducing in Eq.(\ref{FES}) the leading dissipative
term, $-\epsilon \rho(\vec{x},t)$, and neglecting higher order
corrections, the resulting set of equations becomes
\begin{equation}
\left\{
\begin{array}{lll}
  \partial_t \rho(\vec{x},t) &=& a\rho-b\rho^{2}+\omega\rho E(\vec{x},t)
  +D\nabla^2\rho+\sigma\sqrt{\rho}\eta (\vec{x},t) \\ \partial_{t}
  E(\vec{x},t) &=& D_{E}\nabla^2\rho (\vec{x},t) {\bm{ - \epsilon
      \rho(\vec{x},t)}}
\end{array}
\right. 
\label{dFES}
\end{equation}
which is, obviously, non-conserving owing to the activity-dependent energy
leakage.

Integrating in time the equation for the energy field, the following extra
terms for the activity equation are generated:
\begin{equation}
  \omega \rho [E(\vec{x}, 0) ~+ \int_0^t dt' \nabla^2 \rho(\vec{x},t')
    ~{{ - \epsilon \int_0^t dt' \rho( \vec{x},t')}} ].
\label{newterms}
\end{equation}
The second term, dominant in Eq.(\ref{FES2}), becomes a higher order
correction here, i.e. it is irrelevant in the renormalization group sense
as compared with the third, non-Markovian, term. The last one is well
known to be the leading non-linearity in the {\it dynamical
  percolation} universality class \cite{Dyp}.  From this perspective,
it is no wonder that the critical behavior of some non-conserving SOC
models (e.g.  forest-fires) has been related to (dynamical)
percolation in the literature \cite{perc}.  Such a class, whose full
(one-field) Langevin equation is:
\begin{equation}
  \partial_t \rho(\vec{x},t) = (a \bm{+ \omega E(\vec{x}, 0))} \rho-b\rho^{2}
  \bm{- \epsilon \omega \rho \int_0^t dt' \rho(t')} +
  D\nabla^2\rho+\sigma\sqrt{\rho}\eta(\vec{x},t),
\label{dyp}
\end{equation}
describes the spreading properties of epidemics with immunization, etching of
disordered solids \cite{etching}, and some aspects of spreading in systems
with many absorbing states {\it without a conservation law} (see \cite{IAS}
for more details).  The term $ - \epsilon \omega \rho \int_0^t dt' \rho(t')$
ensues that regions already visited by activity become less prompt to be
active in the future. Owing to this term, Eq.(\ref{dFES}) cannot sustain an
active phase. However, even if it lacks a stable active phase, it exhibits a
spreading phase transition separating a phase in which seeds of activity
propagate indefinitely (in the form of rings of expanding activity;
i.e. defining an ``annular growth'' phase) from an absorbing phase in which
they do not \cite{Dyp} (see the diagram at the right part of
Fig.~\ref{Panels}).

This spreading transition, whose critical dimension is $d_c=6$, is controlled
by the initial state in which the seed of activity is placed; indeed, the
initial energy, $ \omega E(\vec{x},t=0) \rho(\vec{x},t)$, is a mass term in
Eq.(\ref{newterms}).  If the initial value of $E$ is large enough
(i.e. favorable environment), then avalanches tend to propagate, while if it
is small, they do not.  Separating these two regimes there is a critical point
for spreading propagation at some value $E^*(\vec{x},0)$.

In summary, {\it the introduction of a non-vanishing dissipation rate affects
  in a relevant way the critical behavior of self-organizing systems};
depending on the initial condition, the dissipative Eq.(\ref{dFES}) can be in
the propagating/supercritical or in the non-propagating/subcritical phase of a
dynamical percolation phase transition.

\subsection{Full theory: dissipation and loading}
\label{53}

Now, we are in a good position to understand in depth the role of the
loading mechanism in dissipative models of SOC, within the Langevin
framework. To do so, let us complement Eq.(\ref{dFES}) with a specific
prescription on how to change the background energy field between
avalanches, i.e. let us consider a loading rule, as for instance:
\begin{equation}
  E(\vec{x},t=0) \rightarrow E(\vec{x},t=0) + \gamma (E_{max}-E)
\label{repob}
 \end{equation} 
 (where $E_{max}$ is a parameter and $E$ the average energy in the
 system) and a driving rule:
\begin{equation}
  \rho(\vec{x_0},t=0) \rightarrow \delta.
\label{seed}
 \end{equation} 
 Eq.(\ref{repob}) and Eq.(\ref{seed}) define one possible loading
 mechanism for Eq.(\ref{dFES}); other choices are, of course, possible
 (for instance, the loading mechanism could also act ``during''
 avalanches). The results presented in what follows are generic,
 essentially independent of such a choice.

 As shown above, Eq.(\ref{dFES}) can sustain avalanches propagating
 indefinitely (up to system size), provided that the initial energy is
 large enough. Therefore, considering large values of $\gamma$ or of
 $E_{max}$ in Eq.(\ref{repob})), the system becomes supercritical for
 avalanche propagation. Instead, small initial densities lead to
 subcritical propagation.

 To illustrate this and the forthcoming discussion, we have performed
 computer simulations of Eq.(\ref{dFES}) using the parameter values
 specified in the caption of Fig.~\ref{Panels}, and complemented with
 the loading and driving rules Eq.(\ref{repob}) and Eq.(\ref{seed}).
 For the sake of simplicity, we have considered $N$ sites with an {\it
   all-to-all} (mean-field) coupling. To check the robustness of our
 conclusions we have also studied a {\it random-neighbor version},
 obtaining very similar results (not shown). As before, the equation
 has been integrated using the algorithm introduced in \cite{Dornic}.

 For a dissipation parameter $\epsilon=0.1$ and fixing $E_{max}=1.3$,
 we find a critical point for spreading at some value of $\gamma$,
 $\gamma_c \approx 2 \times 10^{-2 }$, which generates, on
 average, an initial energy $E^* \approx 1.277(5)$.  At such a
 critical value, power-laws for avalanche and spreading exponents are
 obtained. In complete analogy with Fig.~\ref{PJ}, for smaller values
 of $\gamma$ (and hence, smaller values of the initial average energy
 density) the avalanche size distribution has an exponential cut-off
 (sub-critical), while for larger values the distribution develops a
 bump for large sizes (super-critical) (results not shown).
 \begin{figure}
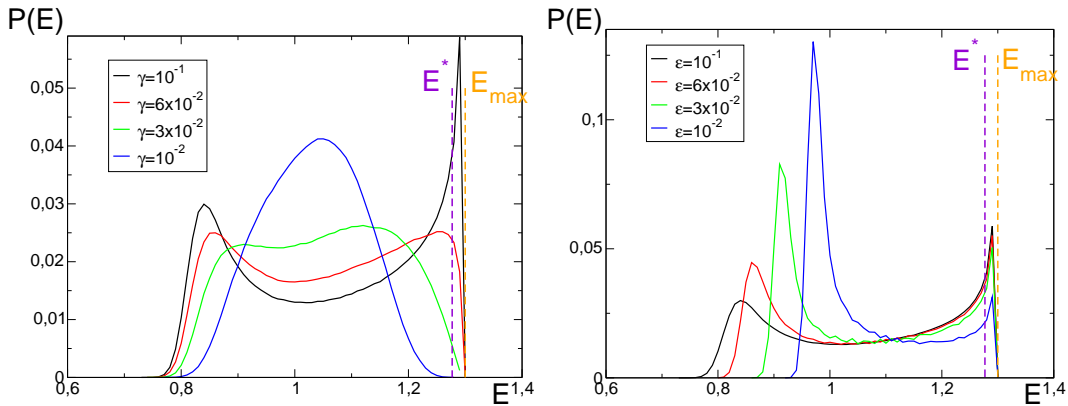
 
\begin{center}
  \includegraphics[height=5.3cm]{CDP1.eps}
  \includegraphics[height=5.3cm]{CDP2.eps}
  \caption{Histograms of $E$, sampled during avalanches, in a numerical
    integration of Eq.(\ref{dFES}) for ({\it left}) $\epsilon=0.1$ and
    different values of the loading parameter $\gamma$ (as soon as $\gamma
    > \gamma_c$ the distribution develops a double-peak shape) and for ({\it
      right}) $\gamma=0.1$ and different values of $\epsilon$. Other parameter
    values are as in Fig.~\ref{Panels}.  }
  \label{histograms}
\end{center}
\end{figure}

We have constructed histograms of the average energy by sampling $E$ during
avalanches in computer simulations (see Fig.~\ref{histograms}).  Typically,
for short times the system is in the right side of the distribution and, as
the avalanche proceeds and dissipation acts, $E$ moves progressively leftward.
This shifting generates a {\it broad distribution of energy values for any
  system size}.  After the avalanche stops, the system is ``loaded'' again
(Eq.(\ref{repob})), a new avalanche is triggered (Eq.(\ref{seed})), and so on.
In this way, the system is kept hovering around the critical point, with a
broad distribution of values, as illustrated in Fig.~\ref{Panels} and
Fig.~\ref{histograms}.

In Fig.~\ref{histograms} (left panel), histograms for $\epsilon=0.1$ and
various values of $\gamma$ are plot.  As long as the loading is sufficiently
strong ($\gamma > \gamma_c \approx 2 \times 10^{-2 } $) the distribution
develops a double-peak structure overlapping with both the propagating and the
subcritical phases. Instead, for smaller values of $\gamma$ ($\gamma=10^{-2}$,
in the figure), the loading is too small, and the histogram overlaps only with
the subcritical phase.  In Fig.~\ref{histograms}, (right panel) histograms for
$\gamma=0.1$ and various values of $\epsilon$ are plot; in all cases, there is
a double-peak structure. Observe that, for large dissipation rates, the system
gets deeper into the absorbing phase.

It is important to remark, that (as illustrated in Fig.~\ref{Panels}) the
dynamics is rather different from its conserving counterpart: while, in the
conserving case, fluctuations around $E_c$ decrease in amplitude with
system-size (Fig.~\ref{Panels}, left-panels), in the non-conserving case the
histograms remain broad, even in the thermodynamical limit (see
Fig.~\ref{Panels}, right panels); i.e.  large variations around the critical
spreading point persist for any system size.

We caution the reader that, within an avalanche, the process is {\it
  not stationary} (energy decreases) and, therefore, the histograms
shown in Fig.~\ref{Panels} and Fig.~\ref{histograms} cannot be
  properly interpreted as probability distribution functions.

\begin{figure}
\begin{center}
  \includegraphics[height=6.0cm]{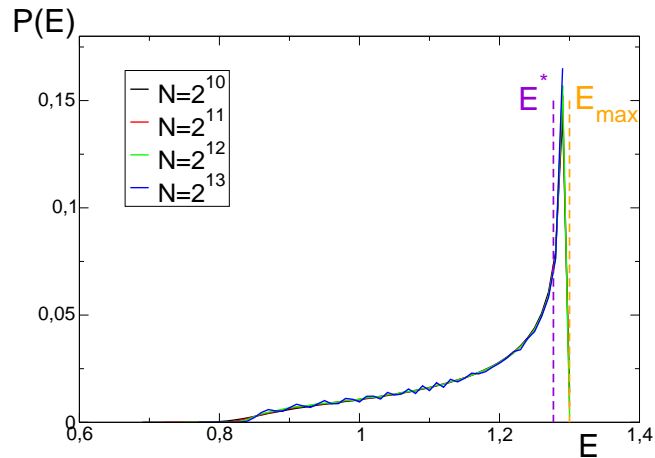}
  \caption{Histograms showing the distribution of initial conditions,
    $E$, for avalanche propagation for various system sizes in
    numerical simulations of Eq.(\ref{dFES}) endowed with the loading
    mechanism Eq.(\ref{repob}) (parameter values are as in
    Fig.\ref{Panels}). Observe that the distribution remains broad in
    the thermodynamic limit.}
  \label{HI}
\end{center}
\end{figure}

  For this reason, we have also constructed {\it stationary} (steady state)
  histograms for {\it (i)} the distribution of the average initial energy for
  avalanches and {\it (ii)} the distribution of $E$ values after avalanches.
  Any of these can be used, as well, to illustrate the differences with the
  conserving case.  For example, Fig.\ref{HI} shows that the background in
  which avalanches are started is generically non-critical, but is broadly
  distributed around the critical point.  Using this information, one can make
  the educated guess that the associated avalanche size distribution (or any
  other quantity measured for avalanches/spreading), with such a distribution
  of initial conditions, will be a convolution of different subcritical and
  supercritical curves weighted with the above distribution of initial
  energies.
Let us emphasize the lack of any mechanism tuning the system to
criticality: the energy is initially set to some arbitrary value
(controlled by the parameter $\gamma$).  If and only if the initial
density is {\it fine tuned} to the critical value for spreading of
activity, $E^*(\vec{x},0)$, the system is at the critical point for
avalanche spreading. Otherwise, for larger values of $\gamma$ the
system is initially supercritical, while for smaller values it is
sub-critical.

In conclusion, fine-tuning of the loading mechanism is required to
have critical spreading in non-conserving systems, and it is
controlled by a dynamical percolation critical point.  The important
point to stress is that, {\it even if the initial condition is not
  critical for generic values of $\gamma$ and $E_{max}$, the
  ``hovering-around-the-critical-point'' mechanism}, illustrated in
Fig.~\ref{Panels} (right panels), {\it keeps the dynamics effectively
  not far from criticality, but not at criticality, for a broad range
  of parameter values}. Furthermore, contrarily to the conserving case
(Fig.~\ref{Panels}, left panels), the energy histograms do not tend to
a delta-peak function for large system sizes.  Large excursions into
both the supercritical and the subcritical phases persist in the
thermodynamic limit.

\subsubsection{Revisiting the mean-field Pruessner-Jensen model.}

Using this insight, we can now tackle the open question: how does
$1/\theta_c(N)$ scales at criticality in the model discussed in
Section \ref{zeroth}?

As $1/\theta_c(N)$ is a background energy, it corresponds to a mass term in
the Langevin equation for the activity, Eq.(\ref{newterms}).  Therefore, to
preserve scale-invariance, it needs to be scaled with $N$ as a distance to the
critical point and, therefore:
\begin{equation}
  1/\theta_c(N) \sim \xi^{1/\nu},
\end{equation}
where $\xi$ is the correlation length and $\nu$ the correlation length
exponent, whose mean-field value in the dynamical percolation class is
$\nu=1/2$ \cite{avalanches}.

At the upper critical dimension, $d_c=6$ (where hyper-scaling mean-field
relations are expected to hold), $\xi$ is limited by the system linear size,
$N^{1/d}=N^{1/6}$, where $N$ is the volume (i.e. total number of sites):
\begin{equation}
\xi \sim N^{1/6}.
\end{equation}
\noindent
Putting together these two last equations, we obtain the mean-field
scaling result,
\begin{equation}
1/\theta_c(N) \sim N^{1/\nu d} = N^{1/3}.
\end{equation}
\noindent 
This is in excellent agreement with the empirical result reported in
Eq.(\ref{scaling}). Analogously, the cut-off of the avalanche size
distribution, $s_c$, needs to scale as
\begin{equation}
s_c \sim \xi^{D_f} \sim  N^{D_f/d_c} = N^{2/3},
\end{equation}
where we have used the fractal dimension, $D_f=4$, for mean-field dynamical
percolation \cite{avalanches}.  Again, this prediction is in perfect agreement
with the numerical results Eq.(\ref{scaling2}).

In conclusion, the loading parameter needs to be fine-tuned for each finite
size to have true scale-invariance; its scaling is inherited from a dynamical
percolation critical point.

\subsubsection{Spatially extended systems.}

The numerical results reported in this section correspond, as already
stressed, to an all-to-all (as well as random-neighbors) coupling in
Eq.(\ref{dFES}). On the other hand, qualitatively similar results can
be obtained for spatially extended systems: i.e. a broad distribution
of the spatially-averaged energy, hovering around the critical point.
The main difference is that, obviously, the background energy becomes
heterogeneously distributed in space and, therefore, the situation
becomes much more involved.

Eq.(\ref{dFES}) generates spontaneously regions with higher and with
lower values of $E$, which have different propensities to activity
propagation: there are {\it locally} super-critical and sub-critical
regions.  Patches where avalanches have passed are typically less
likely to propagate new activity owing to the non-Markovian term in
Eq.(\ref{dFES}).  The size distribution of such patches is,
accordingly, inherited from the avalanche size distribution, creating
a complex landscape for further avalanche propagation.  This scenario
is, of course, more complex than in the mean-field one discussed
above, but the essence of the described phenomenology remains
un-altered: the (local) control parameter hovers-around a dynamical
percolation critical point, with fluctuations that do not vanish in
the large system-size limit.

Observe that, in order to tune the system to criticality (as done in
the mean-field case) one would need in this case to define a more
complicated loading mechanism which should get rid of the dynamically
generated heterogeneities, leading to a homogeneous initial-energy
state, tuned exactly to its critical value.  Using the language of
\cite{BJP}, one needs to ``hire a babysitter'' (or a ``gardener''
using the forest-fire terminology \cite{Broker2}) to keep the
spatially extended system sitting (everywhere) at criticality. Once
such an efficient babysitter is at work, the initial condition is
always at the (dynamical percolation) critical point in any
dimension.

From this perspective, our theory provides additional support for the
claim in \cite{Mega_Grassberger} that the critical density of trees in
a forest fire model should coincide with the percolation critical
density, i.e.  in order to observe critical propagation in the
forest-fire model one should tune the initial background (number of
trees) to the corresponding dynamical percolation critical density.

As already pointed out by Grassberger some time ago \cite{Mega_Grassberger},
in the absence of an efficient gardener taking care of local fine-tuning,
partial power-laws are still observed in the {\it forest-fire model}.  This is
due to the existence of patches with different densities of trees, which
appear with a broad spectrum of sizes. Each of such patches lies at a
different distance of the critical point. The convolution of avalanches
propagating in such a variety of initial conditions originates a complex {\it
  pseudo-scaling picture} which, obviously, does not correspond to strict
criticality. A similar picture applies also to the (open-boundaries) OFC
earthquake model \cite{WD}, and to self-organized models of neural activity
(as will be illustrated in a forthcoming paper).

\vspace{0.5cm} Summing up: In this section, we have reviewed the standard
absorbing state phase transition picture of SOC, underlining the special role
played by conservation. Then, we have introduced a bulk dissipation term and
illustrated that the active phase disappears and the universality class of the
involved spreading phase transition is changed from {\it C-DP} to {\it
  dynamical percolation}.  Dissipation needs to be compensated by a loading
mechanism (which controls the initial conditions in which avalanches are
started) to keep the energy balance.  If and only if the loading mechanism is
perfectly fine-tuned to generate a precise initial energy density, true
criticality is observed.  Otherwise, the system just hovers around a dynamical
percolation critical point, with large excursions into the propagating and the
absorbing phases.  Contrarily to the conserving case, such fluctuations do not
disappear in the thermodynamic limit. {\it Strictly speaking}, this mechanism
of self-organization cannot be called critical, we propose to refer to it as
{\it self-organized quasi-criticality} (SOqC).  Last but not least, our
approach provides a way to rationalize the finite size scaling properties of
non-conserving self-organized systems, as earthquake or forest-fire models,
and has allowed us to derive a number of previously unknown scaling relations.
 
\section{Concluding Remarks}
\label{Last}

We have shown, by using different levels of description, that non-conserving
models of self-organized criticality, as earthquake and fire-forest models,
are not truly critical.

{\it First}, we have studied a simple mean-field theory, based on an energy
balance equation, for a non-conserving model introduced by Pruessner and
Jensen. It permits us to illustrate that, even if one can construct
non-conserving models that seem critical in the thermodynamic limit, there is
no systematic way to have a coherent finite-size scaling description of them:
a precise {\it fine tuning is required} for any finite size to observe
criticality and to approach the thermodynamic limit in a scale-invariant way.

{\it Second}, we have revisited the mapping of high-dimensional (mean-field)
avalanching systems into a self-organized branching process, introduced by
Zapperi {\it et al.}  some years ago.  The underlying idea is that in high
dimensions avalanches do not visit twice a given site and they can be
described as a branching process, whose branching probability depends on the
energy background.  This allows us to write an evolution equation for the
branching probability for models with slow driving and dissipation.  While in
the conserving case the branching probability converges to its critical value,
in the presence of bulk-dissipation the convergence is towards a {\it
  subcritical} point. Introducing a loading mechanism (which mimics the growth
of new trees in forest fire automata or the continuous building-up of stress
in earthquake models), we have shown that {\it the fixed-point towards which
  the system self-organizes can be either critical, subcritical, or
  supercritical}.  Contrarily to previous claims, a fine-tuning of the loading
mechanism is required to reach criticality within this approach.

{\it Third}, we have introduced a full stochastic description of SOC systems
in terms of Langevin equations.  We have reviewed how conserving systems
self-organize to a critical point with well known critical exponents (in the
conserved-directed-percolation universality class).  Instead, for dissipative
systems with a loading term, the dynamics is found to {\it hover around} a
critical point, with large excursions into the absorbing and the propagating
phases, which do not disappear in the large system size limit.  Therefore,
such systems are not generically critical. Still, some traits of the
underlying dynamical percolation critical point can be observed, depending on
the loading parameter and system-size, i.e depending on how large the
excursions into the absorbing/propagating phase are.

\vspace{0.5cm} All these three approaches provide overwhelming evidence that
conserving-dynamics is a necessary condition to observe self-organization to
criticality. Instead, in non-conserving (dissipative) systems equipped with a
loading mechanism, a fine tuning of a loading parameter is required to have
the system sitting at a critical point. Otherwise, the system just {\it hovers
  around} a critical point, with broadly distributed fluctuations which do not
dissappear in the thermodynamic limit: for a broad range of parameters,
non-conserving systems can be fluctuating in the vicinity of a critical point,
but not {\it at} the critical point. We propose to call this: {\it
  self-organized quasi-criticality}..

This conclusion extends to some other non-conserving models of SOC as those
for synchronization of integrate-and-fire oscillators described in
\cite{Bottani,Albert1,Albert2}, the model in \cite{Paolo}, or the model of
neural avalanches in \cite{Levina} (as will be explained in a separate
publication).\\

{\it Is it sensible to refer to self-organized dissipative systems as
  ``critical''?}

The answer to this question is mostly a matter of taste, and depends on what
one wants to define as criticality.  Being strict and calling critical only to
systems sitting at a critical point (allowing at most for fluctuations that
vanish in the thermodynamic limit), then {\it non-conserving systems are not
  truly critical models}.

Being more permissive, one could accept, in principle, the term ``critical''
to refer to systems hovering around a critical point (with persistent
excursions into the subcritical and supercritical regimes), which exhibit
``dirty scaling''.

In order to avoid miss-understandings and miss-conceptions, we strongly favor
the use of an alternative terminology like ``almost criticality''
\cite{Prado}, ``pseudo-criticality'', or, as we said, {\bf self-organized
  quasi-criticality} (SOqC) to refer to non-conserving self-organized systems,
and suggest to restrict the term ``critical'' for truly scale-invariant
systems.

Actually, in many cases, strict criticality might not be required to explain
empirical (truncated) power-laws distributions observed in the real world and,
therefore, ``self-organized quasi-criticality'' remains a useful concept,
despite of the somehow inappropriate and certainly confusing use (and abuse)
of the word ``critical'' in the literature.  Under this light, one could
reconsider the empirical observations discussed in the introduction, as well
as similar ones for which power law distributions have been reported. A
critical inspection of them reveals in many cases that empirical data are
better described by truncated power-laws rather than by pure power-laws
\cite{Newman}.

\vspace{1cm}

\section*{Appendix A: Basic Models in a nutshell}

For the sake of completeness, in this Appendix we present some of the basic
toy models of self-organized criticality. All of them are defined in a
$2$-dimensional lattice (generalizations to $d$-dimensions, to random-neighbor
or all-to-all couplings are straightforward).

\subsection*{\bf {Sandpile-like Models}}

Consider a (height or ``energy'') variable $z(i,j)$, which takes
integer (non-negative) values at each site $(i,j)$ of a two-dimensional
lattice.  The ingredients of sandpiles, ricepiles, and related models
can be sketched as:

\begin{itemize}
\item {\it Slow Driving}: A small input of energy is externally introduced
  into the system (grains are dropped), usually at a single site:
  $z(i,j) \rightarrow z(i,j) +1$.

\item {\it Activation}: A site receiving energy, stores it until a given {\bf
  threshold}, $z_{thr}$, is exceeded and the site is declared {\it active};
  otherwise nothing happens and the system is driven again.

\item {\it Relaxation (or ``toppling'')}: Each active site redistributes (all
  or a fraction of) its accumulated energy among its neighbors.  A relaxation
  event can trigger a chain reaction or {\it avalanche} by activating its
  neighbors and so forth.

\item {\it Boundary Dissipation}: When redistribution events reach the (open)
  boundaries of the system, energy is dropped off.

\item {\it Iteration}: When activity has ceased, the avalanche stops, and a
  new external input is added. The driving/dissipation cycle is iterated until
  a statistically stationary state is reached.

\end{itemize}
For each specific model, the relaxation rules and some other details can
change, giving rise to a zoo of models. We enumerate some of the more commonly
studied ones:

{\bf The Deterministic Bak-Tang-Wiesenfeld Sandpile Model:} The threshold is
fixed to $z_{c}=4$ ($2d$ in $d$-dimensions). Active sites relax according to:

\begin{equation}
\left\lbrace
\begin{array}{rlr}
  z(i,j) \rightarrow& z(i,j) - 4& \\
  z(i',j') \rightarrow& z(i',j') + 1& \;\;\; 
  \textnormal{ for all } (i',j') \textnormal{ n.n. of }(i,j),
\end{array}
\right.
\label{BTW_toppling_2d}
\end{equation}
that is, an active site is emptied and its grains are {\it deterministically}
redistributed amongst its nearest neighbors. The energy at sites out of the
system is fixed to $z=0$, enforcing boundary dissipation when boundary sites
topple. Avalanches measured in such a steady state were originally claimed to
be critical \cite{BTW}. Later work showed that actually, owing to the {\it
  deterministic} nature of the model and, as a consequence, to the existence
of many toppling invariants and breakdown of ergodicity \cite{ergo} the system
is not truly critical but exhibits anomalous multi-scaling
\cite{Stella}. Other authors (see, for instance, \cite{Priezzhev}) suggested
that avalanches do not obey any type of scaling whatsoever.  To avoid the
pathologies associated with deterministic rules, we focus all along this paper
on {\it stochastic} models.

{\bf The Stochastic Manna Model:} The dynamics of the Manna model is similar
to the deterministic BTW, but the threshold is fixed to $z_{c}=2$ in any
dimension and stochasticity is introduced in the redistribution rule
\cite{Manna}:
\begin{equation}
  \left\lbrace
\begin{array}{rlr}
  z(i,j) \rightarrow& z(i,j) - 2& \\
  z(i',j') \rightarrow& z(i',j') + 1& \;\;\; 
  \textnormal{ for $2$ randomñly chosen } (i',j') \textnormal{ n.n. of }(i,j).
\end{array}\right.
\label{Manna_toppling_2d}
\end{equation}
This is the ``inclusive'' version of the model. In its ``exclusive'' version,
the two grains are forced to go to different neighbors. Both of these versions
self-organize the model to a critical point in the C-DP (or ``Manna'') class
in any dimension \cite{Manna,Romu,Lubeck}.

{\bf The Stochastic Oslo Model:} This (ricepile) model has annealed random
thresholds at each site: every time a site topples, a new threshold
$z_{thr}(i,j)=2,3$ is randomly chosen with equal probabilities \cite{Oslo}.
Redistribution of energy units (grains) is done as in the Manna model, in a
stochastic fashion. These rules lead to self-organization to a critical point
in the C-DP universality class with rather clean scaling
\cite{Oslo,Romu,Lubeck}.

\subsection*{{\bf The Olami-Feder-Christensen Earthquake Model}}

A continuous ``force'' or ``energy'' $F(i,j)$, randomly distributed between
$0$ and a threshold value, $F_{thr}$, is initially assigned to each site
$(i,j)$ of a two-dimensional lattice. During the driving step, the site with
maximum force, $F_{i,j}^{max}$, is identified, and the force of all sites is
increased by $F_{thr}-F_{i,j}^{max}$, creating (at least) one seed of activity
(this is equivalent to advancing all sites at a fixed constant velocity until
one of them, the maximum, reaches the threshold). Contrarily to sandpile
models, this driving affects all sites. An active site relaxes according to:
\begin{equation}
\left\lbrace
\begin{array}{ll}
  F(i,j) \rightarrow& 0 \\
  F(i',j') \rightarrow& F(i',j') + \alpha F(i,j) 
  \;\;\;\forall\; (i',j') \;\;\;\textnormal{n.n. of }(i,j),
\end{array}\right.
\label{OFC_toppling}
\end{equation}
where $\alpha \leq 1/4$ is a {\it bulk-dissipation parameter}.  The model is
conserving only for $\alpha=1/4$.  The force at sites out of the system is
fixed to $F=0$, entailing boundary dissipation.  The stationary state of such
a system was claimed to be critical for a broad range of values of $\alpha$
\cite{OFC}, but recent analyses disprove such a claim (see for instance
\cite{QGrassberger,WD} and Section \ref{Intro}).

\subsection*{{\bf The Drossel-Schwabl Forest Fire Model}}

Sites of a two-dimensional lattice can be either empty ($z=0$), occupied by a
tree ($z=1$), or by a tree on fire ($z=2$) \cite{FFM2}. The dynamics proceeds
as follows:
\begin{itemize}
\item At every {\it empty} site, a tree grows with probability $p$,
  and the site becomes ``occupied'': $z=0 \rightarrow z=1$

\item Initial spark: A tree not surrounded by any fire becomes a ``burning''
  tree (e.g. lightening) with probability $f$: $z=1 \rightarrow z=2$

\item A burning tree sets on fire all its nearest neighbors, and it becomes
  empty: $z=2 \rightarrow z=0$

\item The ``fire avalanche'' proceeds burning all trees in contact with fires.

\item When fire ceases the dynamical processes is re-started.
\end{itemize}
The relevant parameter is $f/p$ which sometimes is called $1/\theta$
\cite{FFM2_Grassberger}.  In the implementation of the model that we use,
$1/\theta$ is the number of trees grown between two consecutive
ignitions. Despite of the initial claims and various forms of reported
``anomalous scaling'', the most recent studies revealed absence of generic
scale-invariance \cite{FFM2_Jensen,Mega_Grassberger}.

\vspace{1cm}

\section*{Appendix B}

To compute the mean number of grains lost in the bulk in the dissipative
self-organized branching process \cite{SOBP}, let us first consider the number
of off-springs, $\varphi$, not-occupied owing either to absence of branching
or to bulk-dissipation:
\begin{equation}
  \varphi = \sum_{j=1}^{m-1} (2\varsigma_j - {\varsigma_{j+1}}),
  \label{varphi}
\end{equation}
then, the average fraction corresponding to bulk-dissipation is
\begin{equation}
  \langle\kappa(q,n)\rangle=\langle\varphi\rangle
  \dfrac{p\epsilon}{(1-p)+p\epsilon}.
\label{SOBPD_q_noise2}
\end{equation}
Noting that the avalanche size is $s=\sum_{j}\varsigma_{j}$ and:
\begin{equation}
\left\lbrace
\begin{array}{l}
  \msum_{k=0}^{m-1}\varsigma_{k}=\msum_{k=0}^{m}\varsigma_{k}-\varsigma_{m}
  =s-\varsigma_{m}\\ \\ \msum_{k=0}^{m-1}\varsigma_{k+1}=\msum_{l=1}^{m}
\varsigma_{l}=s-1;
\end{array}
\right.
\label{SOBPD_q_noise3b}
\end{equation}
then
% \begin{equation}
 $ \varphi={1+s-2\varsigma_{m}}$.
%\label{SOBPD_q_noise3c}
%\end{equation}
Using the definition of $s$:
\begin{equation}
\begin{array}{c}
  \langle s \rangle=\msum_{k=0}^{m}\langle \varsigma_{k}\rangle=\msum_{k=0}^{m}
  \left(2p(1-\epsilon)\right)^{k}=\dfrac{1
    -\left(2p(1-\epsilon)\right)^{m+1}}{1-2p(1-\epsilon)}.
\end{array}
\label{SOBPD_q_noise3d}
\end{equation}
Plugging Eq.(\ref{SOBPD_q_noise3d}) into the equation for $ \varphi$,
\begin{equation}
  \langle\varphi\rangle=\left[1+\dfrac{1-\left(2p(1-\epsilon)\right)^{m+1}}
    {1-2p(1-\epsilon)}-2\left(2p(1-\epsilon)\right)^{m}\right],
\label{SOBPD_q_noise3e}   
\end{equation}
and, from this and Eq.(\ref{SOBPD_q_noise2}):
\begin{equation}
  \langle\kappa(q,n)\rangle=\left[1+
    \dfrac{1-\left(2p(1-\epsilon)\right)^{m+1}}{1-2p(1-\epsilon)}-
    2\left(2p(1-\epsilon)\right)^{m}\right]\dfrac{p\epsilon}{(1-p)+p\epsilon},
\label{SOBPD_q_noise_def}
\end{equation}
leading, once the continuum limit for $n$ has been taken and
fluctuations included, to Eq.(\ref{Juanico_p_SOBP}).

%%%%%%%%%%%%%%%%%%%%%%%%%%%%%%%%%%%%%%%%%%%%%%%%%%%%%%%%%%%%%%%%
\vspace{0.7cm}

{\bf Acknowledgments:} We acknowledge financial support from the Spanish
Ministerio de Educaci\'on y Ciencia (FIS2005-00791) and Junta de
Andaluc{\'\i}a (FQM-165). We also thank our friends and colleagues I. Dornic,
F. de los Santos, P. Hurtado, P. Garrido, G. Pruessner, P. Grassberger,
R. Pastor-Satorras, H. Chat\'e, M. Alava, S. Zapperi, A. Vespignani,
R. Dickman for critical reading of the manuscript, for useful comments, and/or
enjoyable collaboration in the past.  This paper is dedicated, with
admiration, to Geoff Grinstein, who introduced us to the art of Langevin
equations, and suspected all this to be true long ago.
 \vspace{0.65cm}

{\bf {References}}
  

\begin{thebibliography}{99}
\bibitem{theta} We warn that $1/\theta$ is called $\theta$ in some papers.
\bibitem{details} More specifically, assuming that in the steady state $F$ is
  homogeneously distributed in the interval $[0,1]$ (which is a good
  approximation \cite{Broker}), the average loading per site between two
  consecutive avalanches is of the order of $1/N$, and hence the global
  loading of the whole system is of the order of $1$.  This cannot compensate
  the dissipation of large avalanches, and the system is damned to remain
  subcritical for any non-vanishing dissipation rate.
\bibitem{fixed} Note that, for $p=1/2$, typically there is just one surviving
  path arriving to the $m$-th generation; for it the probability to not
  dissipate until the end is $(1-\epsilon)^m$, while the corresponding
  boundary dissipation can be seen to be $1-(1-\epsilon)^m)$; hence, they sum
  $1$ and cancel the contribution of the starting seed in the balance
  equation.
\bibitem{quenched} The term, $ \omega \rho E(\vec{x}, 0)$ contributes
  to the linear (``mass'') term. Even if this term looks like a
  quenched variable, the local ``mass'' for the activity is not a
  quenched variable at all; the term $ \omega \rho [\int_0^t dt'
  \nabla^2 \rho(t')]$ converts the effective ``mass'' at each site, $a
  + \omega E(\vec{x},t)$, into an annealed variable, evolving with the
  dynamics.
\end{thebibliography}
\end{document}